\documentclass[sigconf, pbalance]{acmart}

\usepackage{amsmath,amsfonts}
\usepackage{algorithm}
\usepackage[noend]{algpseudocode}
\usepackage{graphbox}
\usepackage{dblfloatfix}
\usepackage{textcomp}
\usepackage{siunitx}
\usepackage{subcaption}
\usepackage{tikz}
\usepackage{pgfplots,pgfplotstable}
\usepackage{placeins}

\copyrightyear{2021} 
\acmYear{2021} 
\setcopyright{rightsretained} 
\acmConference[Middleware '21]{22nd International Middleware Conference}{December 6--10, 2021}{Virtual Event, Canada}
\acmBooktitle{22nd International Middleware Conference (Middleware '21), December 6--10, 2021, Virtual Event, Canada}\acmDOI{10.1145/3464298.3493403}
\acmISBN{978-1-4503-8534-3/21/12}

\begin{document}
	
	\title{Implicit Model Specialization through \\DAG-based Decentralized Federated Learning}

\author{Jossekin Beilharz}
\authornote{These authors contributed equally to this research}
\orcid{0000-0002-9970-3835}
\affiliation{%
	\department{Hasso Plattner Institute}
	\institution{University of Potsdam}
	\city{Potsdam}
	\country{Germany}
}
	\affiliation{
		\institution{Charité – Universitätsmedizin Berlin}
		\city{Berlin}
		\country{Germany}
	}
\email{jossekin.beilharz@hpi.de}

\author{Bjarne Pfitzner}
\authornotemark[1]
\orcid{0000-0001-7824-8872}
\affiliation{%
	\department{Hasso Plattner Institute}
	\institution{University of Potsdam}
	\city{Potsdam}
	\country{Germany}
}
\email{bjarne.pfitzner@hpi.de}

\author{Robert Schmid}
\authornotemark[1]
\orcid{0000-0003-1092-9529}
\affiliation{%
	\department{Hasso Plattner Institute}
	\institution{University of Potsdam}
	\city{Potsdam}
	\country{Germany}
}
	\affiliation{
		\institution{Charité – Universitätsmedizin Berlin}
		\city{Berlin}
		\country{Germany}
	}
\email{robert.schmid@hpi.de}

\author{Paul Geppert}
\affiliation{%
	\department{Hasso Plattner Institute}
	\institution{University of Potsdam}
	\city{Potsdam}
	\country{Germany}
}
\email{paul.geppert@hpi.de}

\author{Bert Arnrich}

\orcid{0000-0001-8380-7667}
\affiliation{%
	\department{Hasso Plattner Institute}
	\institution{University of Potsdam}
	\city{Potsdam}
	\country{Germany}
}
\email{bert.arnrich@hpi.de}

\author{Andreas Polze}
\affiliation{%
	\department{Hasso Plattner Institute}
	\institution{University of Potsdam}
	\city{Potsdam}
	\country{Germany}
}
\email{andreas.polze@hpi.de}

  \renewcommand{\shortauthors}{Beilharz, Pfitzner and Schmid, et al.}

\begin{abstract}

Federated learning allows a group of distributed clients to train a common machine learning model on private data.
The exchange of model updates is managed either by a central entity or in a decentralized way, e.g. by a blockchain.
However, the strong generalization across all clients makes these approaches unsuited for non-independent and identically distributed (non-IID) data.

We propose a unified approach to decentralization and personalization in federated learning that is based on a directed acyclic graph (DAG) of model updates.
Instead of training a single global model, clients specialize on their local data while using the model updates from other clients dependent on the similarity of their respective data.
This specialization implicitly emerges from the DAG-based communication and selection of model updates.
Thus, we enable the evolution of specialized models, which focus on a subset of the data and therefore cover non-IID data better than federated learning in a centralized or blockchain-based setup.

To the best of our knowledge, the proposed solution is the first to unite personalization and poisoning robustness in fully decentralized federated learning.
Our evaluation shows that the specialization of models emerges directly from the DAG-based communication of model updates on three different datasets.
Furthermore, we show stable model accuracy and less variance across clients when compared to federated averaging.

\end{abstract}


\maketitle

\section{Introduction} %
\label{sec:introduction}

\begin{figure}[t!]
	\centering
	\vspace{2.72em}
	\includegraphics[clip, trim={0 0cm 0 0},width=0.98\linewidth, page=1]{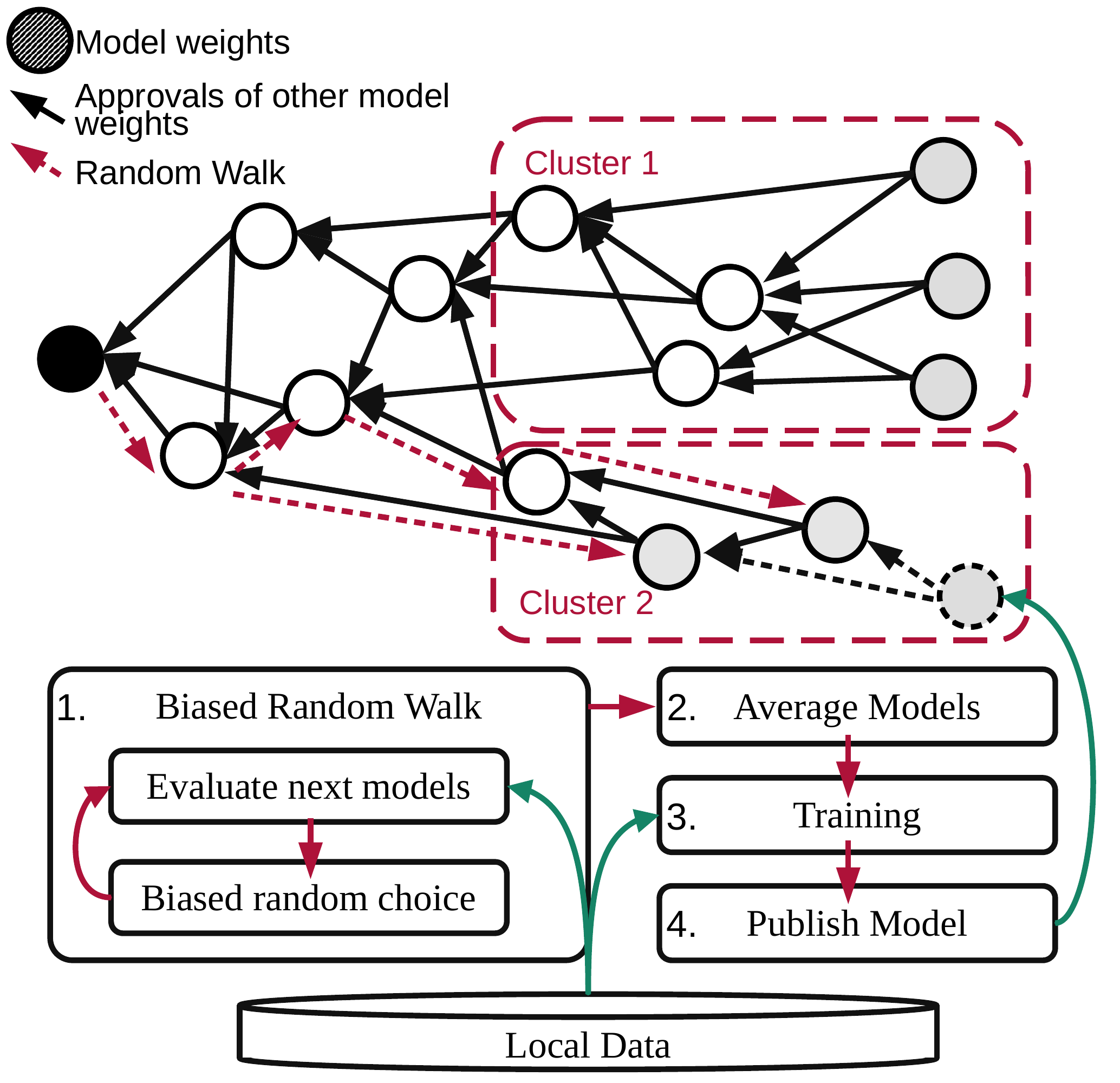}
	\caption{We use a biased random walk through a DAG of model updates to find models that perform well on local data, resulting in clusters emerging in the DAG.}
	\label{fig:fig1}
\end{figure}
Large datasets are often required in order to build powerful and accurate machine learning and especially deep learning models.
This is problematic for many domains such as healthcare, since privacy regulations hinder sharing the sensitive data to create central databases.
Instead, the data has to be used at the site where it was collected, resulting in multiple, sub-optimal, local models.
As a solution, federated learning has been proposed, which allows many widely distributed nodes to train a common machine learning model on local data by exchanging updates to the model, but not the data itself.~\cite{kairouz_advances_2019}
The common model is usually exchanged via a central server, but the communication using decentralized consensus protocols like blockchains has been proposed.~\cite{toyoda_mechanism_2019, qu_decentralized_2020, chai_hierarchical_2020,lu_differentially_2020}
The model updates are stored on the chain and the participants jointly form a consensus by agreeing on the latest changes, thus defining the current global model parameters.

Another common challenge for federated learning and other decentralized learning approaches is the difference in data distributions present for different clients.~\cite{zhao_federated_2018, hsieh_noniid_2020,chen_asynchronous_2020}
For this not independent and identically distributed (non-IID) data, model updates could counteract each other and hinder the training progress.~\cite{zhao_federated_2018}

In this paper, we demonstrate for the first time how the seemingly diverse goals of distributed model training, model personalization as well as robustness against poisoning attacks, can be addressed by a single mechanism that is inspired by distributed ledgers and federated averaging.

Specifically, we propose a fully-decentralized algorithm for solving a federated learning task, utilizing a model selection mechanism that incorporates the performance of other models on the local data:
The so-called \emph{accuracy tip selection} results in implicit model specialization for clusters of clients holding similar data.
Using a synthetic dataset derived from MNIST, we demonstrate how the algorithm creates model convergence while still allowing specializations across different clusters.
Furthermore, we show how malicious clients are isolated within the network, limiting their effects on other participants' models.

Additionally, we discuss means to tune the balance between generalization and specialization of local models as well as measures for the impact of this balancing parameter on cluster formation.
In a quantitative evaluation, we compare the performance of our novel algorithm against centralized federated averaging for two more datasets, namely a dataset consisting of texts from William Shakespeare and Johann Wolfgang von Goethe, as well as CIFAR-100.

In this paper, terms that have different meanings in the context of distributed systems, graph theory, blockchains and federated learning are used.
We use these terms with their following meaning:
When using the term \emph{node}, it is always specified whether a node of the graph or a compute node is meant.
\emph{Transaction} is only used in its meaning in the context of blockchains.
In our approach each node of the directed acyclic graph represents a transaction, thus the terms ``node of the graph'' and ``transaction'' are used to describe different aspects of the same concept.
\emph{Consensus} is only used in its meaning in distributed systems and blockchains.
\emph{Weights} are used in three different contexts, which are clarified in each use: For the trained parameters of the machine learning model we specify them as ``model weights''. ``weights of transactions'' or ``weights of the random walk'' are weights that are used in the random walk as part of the DAG-based consensus. Finally, ``edge weights'' are only used in Section~\ref{sec:approach_measuring} to talk about the modularity of a graph of clients that is different from the DAG that is used for consensus.

The remainder of this paper is structured as follows.
The next section introduces the background to our work, while Section~\ref{sec:relatedwork} discusses concrete examples of related work.
Section~\ref{sec:approach} explains our approach and the implications for model performance, cluster identification, and poisoning robustness in detail.
Section~\ref{sec:evaluation} presents the datasets and models that we used to evaluate our approach, as well as the results of the evaluation itself.
Finally, Section~\ref{sec:conclusion_and_future_work} concludes this paper.

\section{Background} %
\label{sec:background}

This paper builds upon research in the fields of federated learning and decentralized consensus mechanisms.
Combining ideas from these fields, our approach is applied to the problem of multi-task learning and learning with non-IID data.

\subsection{Federated Learning} %
Initially proposed in 2016, federated learning is a novel research area for developing machine learning methods for distributed datasets.~\cite{mcmahan_communicationefficient_2017}
Federated learning relies on a client-server architecture, where the server defines a model to be trained, distributes the model weights to clients, and aggregates new model weights received back from clients.
These are found after training the model for a number of epochs on the private, local client data.
In this iterative fashion, the model is jointly improved until reaching optimal performance.
In contrast to the field of distributed machine learning, the training facilitator does not have any control of, or access to client data.
This brings privacy benefits for collaborators, since their data never leaves their location, but also entails problems with the training process if data is not independent and identically distributed.~\cite{zhao_federated_2018}
Moreover, a key issue related to this is the communication cost, which becomes one of the main hindrances for quick convergence.
Client devices may not always be available or have a bad internet connection in addition to potentially not having fast hardware for machine learning model optimization.

Recent works have suggested that the core goal of federated learning, to jointly develop a single model for all participants, is not ideal and has to be altered.~\cite{jiang_improving_2019}
Instead, federated learning should also consider optimizing for the mean model performance per client, meaning that model personalization is desirable.
A single model might not be able to optimally adapt to the non-IID nature of client datasets.

\subsection{Permissionless Consensus}
Consensus  is a fundamental problem of distributed computing with a large body of research.
Traditionally, consensus research focused on permissioned consensus, where the participating nodes of the consensus are known and authenticated to each other.
Since the publication of Bitcoin~\cite{nakamoto_bitcoin_2008} in 2008, there has been more research in permissionless decentralized consensus schemes.~\cite{buterin_nextgeneration_2014,gilad_algorand_2017,kiayias_ouroboros_2017}
In a blockchain, as introduced by Bitcoin, consensus on the ordering of transactions is reached.
Network participants are selected to propose the next block, often through a proof-of-work (PoW) mechanism.
Participants then select the longest chain of blocks as the correct chain, thus abandoning forks that were introduced by other compute nodes at the same time.~\cite{nakamoto_bitcoin_2008}

There are scalability limits of linear blockchains that lead to congestion in the network.~\cite{huberman_monopoly_2017}
Different proposals have been made to solve these constraints.~\cite{croman_scaling_2016,poon_bitcoin_2016,luu_secure_2016,buterin_casper_2017}
Recent distributed ledger technologies (DLTs), such as Byteball, Spectre, IOTA, Snowflake, and Conflux~\cite{churyumov_byteball_2016,sompolinsky_spectre_2016,sergueipopov_tangle_2017,teamrocket_snowflake_2018,li_scaling_2018}, have replaced the linear blockchain with a DAG.
Similar to a blockchain, the DAG can be used to create consensus in a distributed system.
The main goal of DAG-based DLTs is to allow for higher throughput of the system by allowing multiple concurrent blocks to be added simultaneously.
Thus, while blockchains try to avoid forks and always select one winning chain, these forks are expected in DAG-based consensus mechanisms.
Because forks are common, the question becomes now how the blocks are reintegrated to form the consensus.
For this, each new block selects not only one, but multiple previous blocks to approve.
Thus, over time, all valid blocks will be approved in the DAG.

Because of the potential for higher throughput, DAG-based consensus protocols are often seen as especially suited for use cases that involve large numbers of widely distributed resource-constrained devices like the internet of things, fog, and edge computing.
\pagebreak[3]
\subsection{Multi-Task Learning and Non-IID Data} %
The domain of multi-task learning deals with solving multiple related tasks simultaneously by transferring knowledge between them during the training process.~\cite{caruana_multitask_1997}
For convex optimization problems such as training a linear or logistic regression model the challenge of multi-task learning often lies in finding a similarity matrix between tasks which is incorporated in the global optimization formulation.~\cite{zhang_convex_2010}
An extension to multi-task learning called clustered multi-task learning further has the premise that some of the tasks are more related than others, forming clusters which all have similar model parameters.~\cite{zhou_clustered_2011,jacob_clustered_2008}

For deep learning model architectures, multi-task learning is usually modeled as soft and hard parameter sharing between tasks.~\cite{ruder_overview_2017}
In the former case, multiple models are learned, one per task, but the weights from different models influence each other, for instance by bounding the distance between them.
The latter type re-uses some of the model layers completely, splitting the model at some point to end up with multiple prediction layers for the various tasks.

Federated learning with non-IID data can also be seen as a multi-task learning problem with the clients forming the different tasks.~\cite{smith_federated_2017,corinzia_variational_2019}
As simplification, clients can be clustered together if their datasets are similar and thus also their training task will be similar.
Federated learning with cluster specialization is thus comparable to multi-task learning with soft parameter sharing.

\section{Related Work} %
\label{sec:relatedwork}

The related work falls into two areas: decentralized federated learning, that is federated learning without a central server, and specialization in federated learning.
Research in the area of decentralized federated learning relies on peer-to-peer networks to distribute the learning progress.
One body of research focuses on the usage of blockchain architectures to create a global consensus on the learned model, while another, called \emph{gossip learning}, relies on the participants themselves to merge models received from peers.
While there is a significant interest in specialization or personalization in federated learning recently, this still is an emerging field of research.

\subsection{Decentralized Federated Learning with DLTs}
Many related works investigate the use of blockchains to communicate model updates in federated learning.~\cite{shayan_biscotti_2018, chen_when_2018, wang_inferring_2018, kim_blockchainbased_2019,kim_blockchained_2019,weng_deepchain_2019}

In these works, the gradient updates are inserted into the distributed ledger.
The current network consensus on the global model is then defined by the model contained in the latest transaction that has a sufficiently high probability of being part of the longest chain of blocks.

Different manifestations of this general architecture have been proposed for different domains, such as health~\cite{rahman_secure_2020, kumar_blockchainfederatedlearning_2020}, smart home~\cite{zhao_privacypreserving_2020} or railway operation~\cite{hua_blockchainbased_2020}.

For urban mobile networks, Lu et~al.~\cite{lu_differentially_2020} address privacy and security concerns for federated learning.
They propose a federated learning scheme where all nodes publish model updates and the corresponding mean absolute error.
Interesting for our work is the use of a directed acyclic trust graph by Lu et~al. to mitigate the risk of malicious nodes publishing wrong mean absolute errors, which would lead to biased averaging in their system.
Schmid et~al.~\cite{schmid_tangle_2020} describe how a \emph{learning tangle}, similar to the IOTA ledger, can be adapted for a decentralized, asynchronous implementation of federated averaging.

One topic in the research on federated learning in open networks is the question of fairness.
That is, how to ensure the participants only benefit from the common machine learning model as much as they contributed to it.
DeepChain~\cite{weng_deepchain_2019} aims to guarantee fairness by a monetary incentive mechanism.
Participants need to deposit monetary value, which is distributed to the other participants if a participant is found to be dishonest.

The FPPDL framework~\cite{lyu_fair_2020} transmits differentially private artificial samples and encrypted model updates via the blockchain.
To ensure fairness, the mutual evaluation mechanism is based on points that can be earned by sharing model updates and then traded for the updates of other participants.
The usage of local models instead of a global one is particularly relevant for this paper.
However, in FPPDL these local models are only meant to restrict the model performance to be in line with the local contribution.

\subsection{Decentralized Gossip Learning}
\emph{Gossip learning} algorithms~\cite{ormandi_gossip_2013,hegedus_gossip_2019,hegedus_decentralized_2021,dinani_gossip_2021,vanhaesebrouck_decentralized_2017,hardy_gossiping_2018,blot_distributed_2019,giaretta_gossip_2019} also utilize a peer-to-peer network, but while blockchain approaches store learning progress on the ledger, this decentralized learning paradigm lets the participants deal with the way they include new information from their neighbors into their model.
Clients periodically send out their current model parameters to a (randomly) selected peer.
Upon reception of new parameters, a client merges their own model and the new one, for instance by taking the average, and updates the resulting model with their local data.
Gossip learning research has investigated different sampling and merging algorithms to improve accuracy~\cite{hegedus_gossip_2019,danner_token_2018,dinani_gossip_2021,vanhaesebrouck_decentralized_2017}, different compression schemes to improve the communication efficiency~\cite{hu_decentralized_2019}, as well as different models such as GANs~\cite{hardy_gossiping_2018}.

Heged\H{u}s et~al.~\cite{hegedus_decentralized_2021} evaluated gossip learning against regular federated learning and found that for IID scenarios both approaches reach similar model performances.
If the data is non-IID, however, gossip learning can struggle to converge quickly due to the lack of a central component.
Looking at the communication efficiency, the authors claim that even though gossip learning requires more network traffic due to the peer-to-peer nature rather than having a single point of contact for all clients, the difference in convergence speed is relatively modest.

Some gossip learning research specifically investigates algorithms for creating personalized models.
For instance, Vanhaesebrouck et~al.~\cite{vanhaesebrouck_decentralized_2017} assume different training objectives per participant and propose two algorithms for incorporating knowledge from other people's models.
Dinani et~al.~\cite{dinani_gossip_2021} put their focus on a dynamic network structure by example of a vehicular ad hoc network and incorporate the other models considering their marginal utility.

Compared to our approach, specialization for gossip learning can be reliant on the availability of related peers in the network which could slow down convergence speed~\cite{giaretta_gossip_2019}.
Moreover, many gossip learning algorithms do not consider robustness against poisoning attacks.
Even though recent work found, that peer-2-peer approaches may be more resilient than federated learning in real-world scenarios with many clients, they also state that more research is required to further investigate this topic~\cite{tomsett_model_2019}.

\subsection{Specialization in Federated Learning}

One approach to allow for personalization in federated learning is to represent different client characteristics within a single global model.
\texttt{FedProx} \cite{liFederatedOptimizationHeterogeneous2020} generalizes and reparameterizes \texttt{FedAvg} to better account for both non-IID data distributions as well as stragglers that only submit partially trained updates into the synchronized averaging rounds.
\texttt{Ditto} \cite{liDittoFairRobust2021} defines fairness and robustness in federated learning and provides an updated optimization objective for the global model.

Other approaches use local models to improve the machine learning performance for individual clients.
The benefit of this personalization has been shown for the language model of a virtual smartphone keyboard.~\cite{wang_federated_2019}
One personalized federated averaging algorithm, called Per-FedAvg~\cite{fallah_personalized_2020} uses model-agnostic meta-learning~\cite{finn_modelagnostic_2017} methods to find an initial shared model that users can adapt to their local data with comparably little training.
Mansour et~al.~\cite{mansour_three_2020} propose three different approaches for the personalization of models: user clustering, data interpolation, and model interpolation.
The user clustering is especially relevant for this paper.
As part of this approach, hypothesis-based clustering is proposed to incorporate the machine learning task into the clustering.
The idea of democratized learning~\cite{nguyen_distributed_2020,nguyen_selforganizing_2020} aims to provide flexible distributed learning systems in which learning agents self-organize in a hierarchical structure to perform learning tasks together.
Adaptive personalized learning~\cite{deng_adaptive_2020} proposes the mixing of local and global models to form a personalized model that performs better than either the local or the global one.

\section{Approach}      %
\label{sec:approach}

In our approach of a Specializing DAG, the training runs in four steps for each client, as described in Figure~\ref{fig:fig1}: the biased random walk (1) selects two tips in the DAG, the models of these two tips are then averaged (2), the averaged model is improved by training (3) it on local data and, if the training improved the model, published (4).

This section describes our approach in detail. We discuss the mechanics of our DAG-based decentralized federated learning in two rounds: Section~\ref{sec:approach_dag} explains the fundamental consensus mechanics applied to federated learning and Section~\ref{sec:approach_specialization} discusses in detail how the accuracy of foreign machine learning model updates can be incorporated. We then go on to  explore its ability for implicit specialization and finally discuss implications for the robustness against poisoning attacks.

\begin{figure}[h!]
  \centering
  \includegraphics[width=1\linewidth, page=1, trim={0 1cm 0 .5cm}]{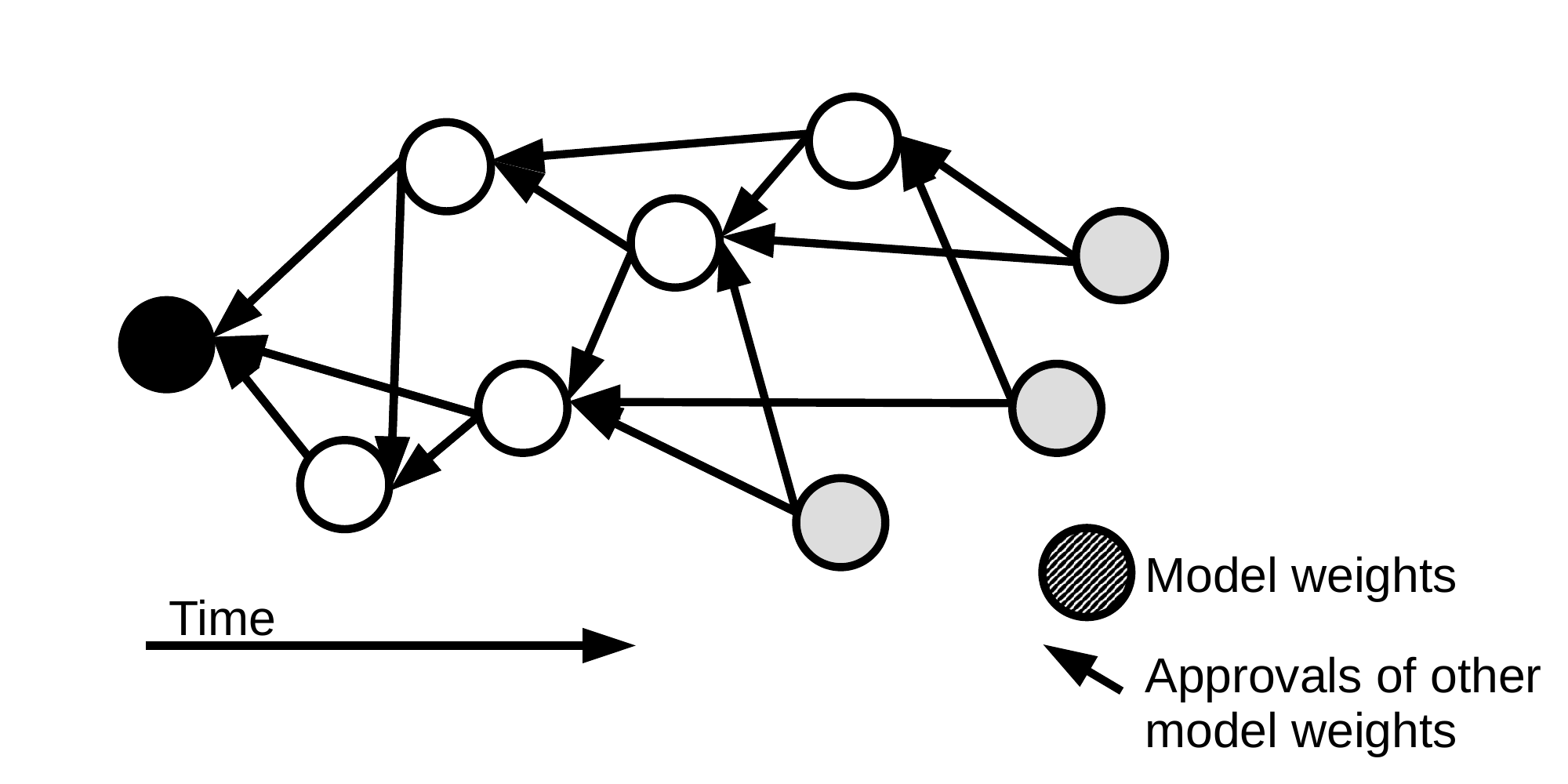}
  \caption{Communicating model updates in federated learning through a DAG: The nodes in the graph are model weight updates and the edges connect a weight update to the two other weight updates that were used as a basis for its training. Tips of the DAG (gray) are updates that didn't receive any approvals yet.}
  \label{fig:dag}
\end{figure}

\subsection{A DAG of Machine Learning Model Updates}
\label{sec:approach_dag}
We propose using a DAG of model updates for decentralized learning.
Specifically, updated model weights are published as nodes in this DAG, while edges represent approvals of previous models that have been the basis of the current one, as illustrated in Figure~\ref{fig:dag}.

Our DAG-based consensus is based on the approach of Popov~\cite{sergueipopov_tangle_2017}, altered in a few key ways to adapt it to our use case of decentralized learning with implicit specialization.
To publish a new model, a client averages two previously proposed models and trains the resulting averaged model on the local data.
The new model update is then published on the DAG as approving the two models it was derived from.
Clients only publish their model update if the training resulted in a model that performs better on the test data than the current consensus model.

Thus, compared to traditional federated learning, we propose to remove the central entity that averages the model weights and replace it with a decentralized consensus mechanism.
Unlike many previous works on federated learning with decentralized models that communicate the model updates via a blockchain, using a DAG for the communication has two benefits.
Firstly, the DAG allows for better scalability: multiple transactions can be proposed at the same time and still be reconciled because the newer transactions approve more than one other transaction. %
Secondly, and more importantly, the structure of a DAG allows for more flexibility in the model communication which is harnessed to create an implicitly specializing system as discussed in the next section.

Our use of the DAG results in a few important differences to traditional DAG-based consensus.
In DAG-based consensus mechanisms as used for cryptocurrencies, each transaction can be checked for consistency with other transactions.
In our DAG of models, the transactions don't fall into the absolute categories 'valid' or 'not valid'.
Rather, the quality of the model is a relative measure.
A further difference in semantics of the DAG is that this quality of a model is dependent on the data the model should be applied to.
When applied to federated learning with strong non-IID data, the quality of a model is very different for each client with its local data.

\subsection{Enabling Implicit Specialization through Accuracy-Aware Tip Selection} %
\label{sec:approach_specialization}

In DAG-based consensus mechanisms, one important aspect is the tip selection algorithm - the algorithm that selects the tips that should be approved when publishing the next transaction.
Schmid~et~al.~\cite{schmid_tangle_2020} showed the general applicability of using DAG-based consensus for decentralized federated learning.

\begin{figure}[h]
  \centering
  \includegraphics[width=1\linewidth, page=2, trim={0 1cm 0 .5cm}]{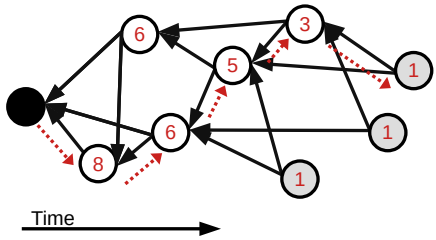}
  \caption{In DAG-based consensus schemes, traditionally weights of transactions are calculated by counting the number of approving transactions (also considering all transactions as self-approving). Thus, the weights are a global property of the DAG itself. The dashed red arrows show a random walk that always chooses the highest weight. }
  \label{fig:walk}
\end{figure}

In this paper, we change this fundamental part of the consensus algorithm by integrating the model performance on the local data as bias of the tip selection algorithm.
The tip selection algorithm is a random walk through the DAG in the opposite direction of the approvals. %
Traditionally, the random walk is biased by assigning each transaction a weight proportional to the size of the subgraph that spans behind it, as illustrated in Figure~\ref{fig:walk}.

We change the bias of the random walk to be specific to a participating client.
In each step during the walk, all potential next models, i.e. those reachable by taking one step in the DAG, are evaluated on the local test data, as shown in Algorithm~\ref{alg:walk}.

\begin{algorithm}
\caption{Random Walk of the Specializing DAG}
\label{alg:walk}
\begin{algorithmic}
	\Procedure{RandomWalk}{Node $ n$}
		\State $\textit{children} \gets$ GetChildren($n$)
		\State initialize \textit{accuracies}
		\For{each $child$ in $children$}
			\State $\textit{accuracies}_\textit{child} \gets$ \Call{EvaluateOnLocalData}{\textit{child}}
		\EndFor
		\State $\textit{nextNode} \gets$ \Call{WeightedChoice}{\textit{accuracies, children}}
		\State \Call{RandomWalk}{\textit{nextNode}}
	\EndProcedure
\end{algorithmic}
\end{algorithm}

The \textsc{WeightedChoice} chooses from these models randomly, weighted by the accuracies of the children on the local data.
These accuracies are normalized by subtracting the maximum accuracy (see Eq.~\ref{eq1}), resulting in negative normalized values.
The weight between 0 and 1 is then calculated by taking the natural exponential of the normalized values scaled by a parameter $\alpha$ (see Eq.~\ref{eq2}).

\begin{align}
\mathit{normalized} &= \mathit{accuracy} - \max(\mathit{accuracies}) \label{eq1}\\
\mathit{weight} &= \exp({\mathit{normalized} \times \alpha}) \label{eq2}
\end{align}

The amount of randomness in the walk can be determined by the $\alpha$ parameter, where higher values result in larger differences between the weights and thus less randomness and more determinism.
Smaller values for $\alpha$, on the other hand, lead to converging weights and thus more randomness.
The expected differences in accuracies between models is dependent on the machine learning problem our approach is applied to, as well as the hyperparameters such as learning rate, batch size, and local epochs.
In order to allow for good specialization even with small changes in accuracy between models and good generalization even with large differences, we include the spread of accuracies $\max(\mathit{accuracies}) - \min(\mathit{accuracies})$ in each step as part of an altered normalized accuracy $\mathit{normalized^*}$:
\begin{equation}
\mathit{normalized^*} = \cfrac
  {\mathit{accuracy} - \max(\mathit{accuracies})}
  {\max(\mathit{accuracies}) - \min(\mathit{accuracies})}
\end{equation}
We show the superiority of this altered normalization for certain values of $\alpha$ in our evaluation in Section~\ref{sec:evaluation}.

This change in the tip selection of the consensus algorithm fundamentally changes the goal of the algorithm: from solely creating a consensus between the distributed clients on a central model to striking a balance between the generalization and specialization of the multiple client-local models.

\begin{figure}[h]
  \centering
  \includegraphics[width=1\linewidth, page=3]{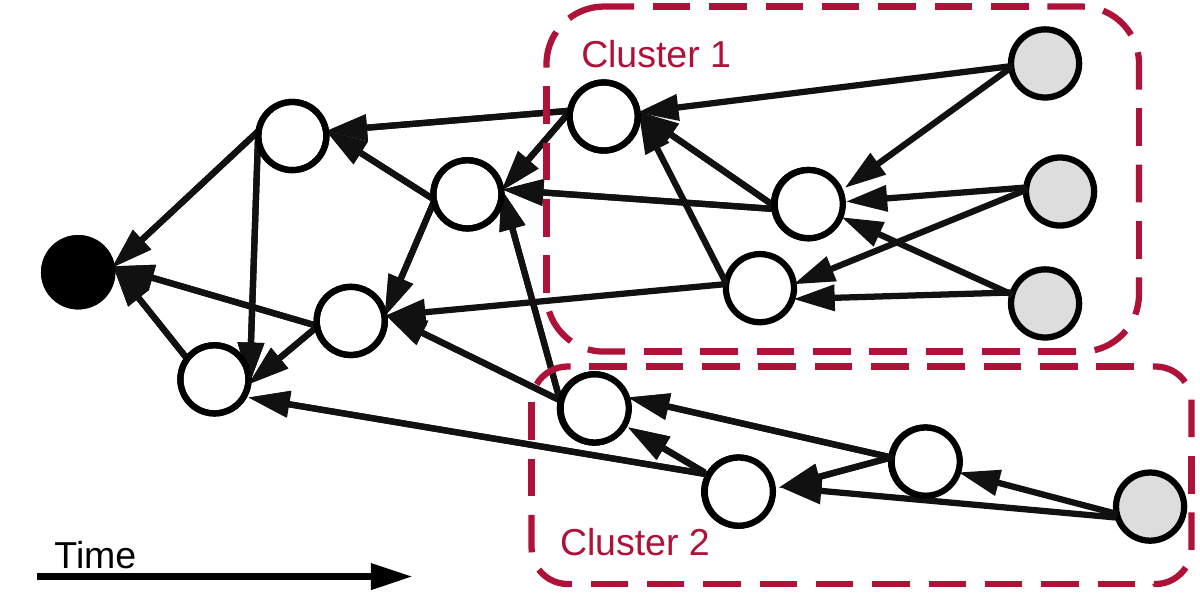}
  \caption{Tip selection using random walks biased by the model performance on local data leads to specialization and clustering in the directed acyclic graph.}
  \label{fig:specialization}
\end{figure}

\subsection{Measuring Implicit Specialization} %
\label{sec:approach_measuring}

As described in the previous sections,
the clients participating in the network implicitly form communities through mutual approvals of each other's transactions.
Since the borders and memberships of these client communities are not directly expressed by the DAG, this section will introduce derived metrics for quantifying the degree of implicit specialization in the network.

The illustration in Figure~\ref{fig:specialization} suggests that clusters in the DAG are visually easy to identify.
However, in visualizations of actual experiment runs with many more model updates, this is not the case, especially since the randomness in the tip selection leads to frequent updates that connect two otherwise disjoint clusters.
This is also the reason why finding the minimum cut within the DAG is not helpful in identifying large subgraphs that can be regarded as clusters.

Since in our experiments the set of participating clients is fixed and known in advance, we instead use a derived graph of clients $G_{\mathit{clients}}$ to identify communities:
In this graph, the edge weight between two clients $c_a$ and $c_b$ is determined by the number of transactions that were published by $c_a$ and directly approve a transaction of $c_b$ or vice versa.

The \emph{modularity} $m \in [-\frac{1}{2}, 1]$ is a measure for the existence of meaningful communities within a graph.
Specifically, given a partitioning of the clients in $G_{\mathit{clients}}$, it expresses the difference between the accumulated edge weights within the partitions and the expected edge weights if they were randomly distributed among all clients in the graph. \cite{newman_finding_2004,newman_modularity_2006}
To obtain a fast approximation of the optimal graph partitioning achieving the highest modularity, we use the \emph{Louvain} algorithm \cite{blondel_fast_2008}.

Since the accuracy-biased tip selection consistently selects models created by clients with similar data characteristics, the modularity of $G_{\mathit{clients}}$ should be positive for every DAG of model updates.
Furthermore, once all clients have participated in training at least once, the edge weights within their communities are expected to continuously increase.
Accordingly, the modularity should eventually converge to 1 during the course of our experiments.

As an additional measure of the specialization quality, we compare the clusters obtained by the Louvain algorithm with the clustering of clients that is known in advance.
The \emph{misclassification} fraction describes how many clients end up in a cluster where the relative majority of clients belongs to a different cluster according to the input labels.

The ability to identify groups of clients with similar characteristics can have negative implications on data privacy when participating in the network:
Although clients publish their model updates anonymously, small clusters increase the potential for de-anonymization attacks.
Furthermore, if characteristics of a single client in a cluster are known, it may be possible to draw conclusions about private data of other clients that belong to the same cluster.

Section~\ref{sec:evaluation_results} discusses in detail how the previously introduced metrics can be used to optimize the random walk in the DAG.

\subsection{Improved Robustness} %

For their decentralized learning DAG \cite{schmid_tangle_2020}, Schmid~et~al. discussed two types of attacks that can be carried out to degrade the prediction accuracy for other participants in the network:
Submitting random weights as well as weights that were trained using a mislabeled dataset, i.e. one in which labels of two classes are flipped.
We adopt their threat model for this paper.

Submitting manipulated weights as a model update can prevent other nodes from publishing their training results because the final model does not improve, thus compute resources are wasted.
Furthermore, if the malicious updates are dominating the network activity, they can take over the consensus, which leads to a degraded prediction performance for others.

With the accuracy-aware random walk, the effects of randomly generated model updates are effectively limited since their prediction accuracy is close to zero.
Thus, given that attackers can only publish malicious updates at a limited rate, they must now make a compromise between poisoning effects and the probability that their transactions are selected by other clients during the biased random walk.

Furthermore, if attackers are aiming to influence the consensus model accuracy for the entire network, they would likely not use the accuracy-aware tip selection strategy as this would limit the effects of their attack to only a subset of the clients in the network.
For targeted attacks at a cluster of clients, the success probability is increased since fewer transactions are necessary to dominate a subgraph of the DAG.
Accordingly, for the sake of poisoning resistance, is is necessary to limit the degree of specialization by choosing a low value for the specialization parameter $\alpha$.

In the remainder of this paper, we discuss a flipped-label attack scenario in which an attacker is able to manipulate the labels in the dataset of one or many clients, e.g. by installing forged sensing hardware.
This means that attackers do not directly submit transactions into the network and cannot influence the tip selection process of the client.
For the overall network, it is now desirable that (1) other clients remain unaffected from the data manipulation and (2) the affected clients are able to detect the data forgery.

In Section~\ref{sec:evaluation}, we discuss if these objectives can be met using the proposed accuracy-aware random walk.

\section{Evaluation}      %
\label{sec:evaluation}
We evaluated our approach on three datasets using a prototype implementation.

This section discusses the datasets and models used, and shows how the specialization emerges in the DAG for each of the datasets.
Furthermore, we compare the performance of our approach against two other federated learning approaches on different machine learning tasks:
Federated Averaging (\emph{FedAvg}) is the original centralized federated learning process and \emph{FedProx} is a state-of-the-art extension of FedAvg that accounts for non-IID data distributions amongst clients.
Finally, we evaluate the poisoning resistance and scalability of our approach.

The prototype implementation of our approach and simulation used for this evaluation was published online.\footnote{\url{https://github.com/osmhpi/federated-learning-dag}}

\subsection{Datasets}
We used three datasets with different characteristics to show the impact of our approach in different scenarios.
Firstly, we evaluate a Handwriting recognition task on a synthetically clustered version of the FEMNIST dataset, in addition to next character prediction on a new dataset from texts by Shakespeare and Goethe, and finally an image classification task on the CIFAR-100 dataset.
Every client dataset has a train-test-split of 90:10.

\subsubsection{FMNIST-Clustered} %
One of the most commonly used datasets for image classification is the MNIST dataset of 28x28 pixel handwritten digits~\cite{lecun2010mnist} and its extension Extended MNIST (EMNIST) which also includes handwritten letters.
For the federated case, the LEAF project~\cite{caldas_leaf_2019} includes a dataset where the images are associated with the person who wrote the digits/letters (FEMNIST).
To better show the effects of our approach, we opted for synthetically clustering clients by class, i.e. digit, and abandoning the split by author.
Specifically, we constructed three disjoint clusters for classes \{0, 1, 2, 3\}, \{4, 5, 6\}, and \{7, 8, 9\} and assigned an equal number of clients to each cluster.

\subsubsection{Poets} %
Our \emph{poets} dataset is an extension of the \emph{Shakespeare} dataset which is often used as a benchmark for federated learning.
We evaluate the applicability of our approach to the task of next character prediction on this dataset.
Poets combines texts from William Shakespeare and Johann Wolfgang von Goethe.
The Shakespeare subdataset was preprocessed by the LEAF framework~\cite{caldas_leaf_2019}.
In addition, we extracted Goethe's plays from \emph{Project Gutenberg} \cite{projectgutenberg_project_2020} and preprocessed them in the same way as the Shakespeare data.
Both subdatasets were cleaned from clients with less than 1000 samples.
To have an equal split between the number of English and German data samples, we reduced the Shakespeare data to 30\% of its size by random sampling.
We assigned the English and German datasets to separate clusters.

\subsubsection{CIFAR-100} %
As image dataset with a non-synthetic clustering we investigated CIFAR-100 \cite{krizhevsky09learningmultiple}, including 32x32 pixel RGB images of different animals, objects or landscapes, belonging to 100 classes, which are each categorized into one of 20 superclasses.
We use those superclasses as the clusters for our learning approach.
The client data allocation was done using the Pachinko Allocation Method (PAM)~\cite{li_pachinko_2006} based on random draws (without replacement) from symmetric Dirichlet distributions over the superclasses and associated subclasses, as used by the Tensorflow Federated framework.
In our experiments, all clients have both training and test data, which is required for calculating the weights of the random walk.
We manually split each client's data into train and test partitions.
In this dataset, clients possess data from more than one superclass/cluster, meaning there is no clear client-cluster affiliation.
For analysis, we choose the cluster per client to be the most common one in its data, choosing randomly in case of a tie.
Our CIFAR-100 dataset consists of 94 clients,
the number of clients per cluster lies between three and six.

\subsection{Models} %
The models for both the FMNIST-clustered and Poets dataset are based on models from the LEAF framework~\cite{caldas_leaf_2019}.
Prediction of FMNIST-clustered digits is done using a Convolutional Neural Net (CNN) with two ReLu activated convolutional layers with kernel size 5, and 32 and 64 filters, respectively, each followed by a max pooling layer with pool size and stride length 2.
Afterwards, a fully connected layer with 2048 neurons and a ReLu activation function leads to the final fully connected layer with 10 output neurons and softmax activation for prediction.

For the Poets dataset, the LSTM model is fed an embedding of dimension 8, calculated from the 80 character sequence. The input is then fed through LSTM layers with 256 units each. Finally there is a dense output layer for prediction.

The classification model for CIFAR-100 is also a CNN similar to the one for FMNIST-clustered.
After the two convolutional layers which are the same for both datasets, there is a third one with 128 filters, also followed by a max pooling layer.
Finally, the model includes two fully-connected hidden layers and an output layer with 256, 128, and 100 neurons, in order.

The fixed training hyperparameters are shown in Table~\ref{table:hyperparams}.

\begin{table}[h]
	\centering
	\begin{tabular}{lccc}
		& FMNIST- &  & \\
		& clustered & Poets & CIFAR-100 \\
		\hline
		Training rounds & 100 & 100 & 100 \\
		Clients / round & 10 & 10 & 10 \\
		Local epochs & 1 & 1 & 5 \\
		Local batches & 10 & 35 & 45\\
		Batch size & 10 & 10 & 10 \\
		Optimizer & SGD(0.05) & SGD(0.8) & SGD(0.01) \\
	\end{tabular}
	\caption{Hyperparameters chosen for the evaluation. The number of local batches is fixed in order to equalize the number of batches used for training per client in case of an uneven distribution.}
	\label{table:hyperparams}
\end{table}

\subsection{Results}
\label{sec:evaluation_results}

We evaluated our approach with a prototypical Python implementation based on the work in LEAF~\cite{caldas_leaf_2019}.
For simplicity, we simulate the distributed training process in discrete rounds.
We present our findings on three topics: optimizing the random walk by choosing good values for $\alpha$ for the FMNIST-clustered dataset, comparing the overall performance of our approach with federated learning as well as the poisoning robustness of our approach.

\subsubsection{Optimizing the Random Walk}

Section~\ref{sec:approach_specialization} describes how the $\alpha$ parameter controls the randomness involved in the biased random walk through the DAG.
When configuring a decentralized learning task, it is necessary to determine a value of $\alpha$ that strikes a good balance between specialization and generalization for the learning task and non-IID data characteristics at hand.
Specifically, we leverage the $G_{clients}$ graph and the metrics introduced in Section~\ref{sec:approach_measuring}.

There are three criteria that indicate an appropriate choice of $\alpha$:
Firstly, the tip selection should be consistent enough so that in a majority of cases clients approve transactions only from other clients from the same cluster.
This can be observed through the \emph{approval pureness} metric.
Additionally, the \emph{modularity} metric of the $G_{clients}$ graph can show how clusters of clients emerge from the approvals without requiring pre-provided cluster labels.

Moreover, the corresponding partitioning of clients should consist of an appropriate number of partitions.
If these are unreasonably many, $\alpha$ is set too high and the client models likely don't generalize well.
Finally, the model differences between clusters should become more distinct over time, which corresponds to a continuously decreasing misclassification fraction.

In our experiments, the approvals in the DAG show perfect pureness (approval pureness of $1.0$) for $\alpha = 10$ and $\alpha = 100$, while $\alpha = 1$ shows less than half of the model updates approving other model updates from within the same cluster (approval pureness of $0.47$).
Still, the pureness for $\alpha = 1$ is higher than the $0.33$ that would be expected for random approvals with three clusters.

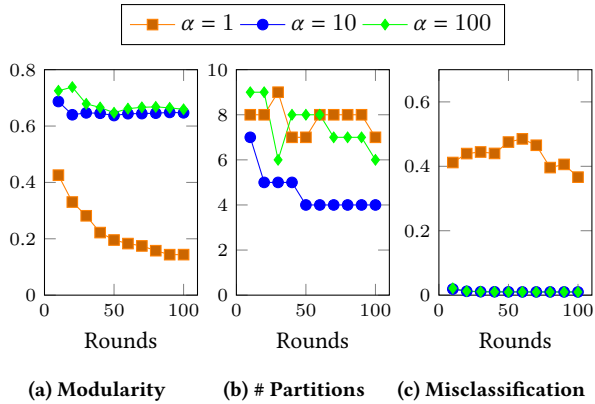
\begin{figure}[h]
  \centering
  \begin{subfigure}[b]{.3\columnwidth}
    \centering
    \begin{tikzpicture}
    \pgfplotstableread{fig/choosing_alpha/lowalpha/modularity_per_round.txt}\low
    \pgfplotstableread{fig/choosing_alpha/alpha10/modularity_per_round.txt}\medium
    \pgfplotstableread{fig/choosing_alpha/highalpha/modularity_per_round.txt}\high
    \pgfplotsset{
        height=3cm,
        width=.8\textwidth,
        compat=1.10,
        tick label style={font=\footnotesize},
        scale only axis,
    }
    \begin{axis}[
        xlabel=Rounds,
        ymin=0, ymax=0.8,
        legend columns=-1,
        legend style={at={(0.5,1.2)}, anchor=west},
        legend cell align={left},
        cycle list name=exotic,
        xmin=0
    ]
        \pgfplotsset{cycle list shift=1}
        \addplot table[x=x,y=y0] {\low};
        \addlegendentry{$\alpha = 1$}
        \addplot [blue, mark=otimes*] table[x=x,y=y0] {\medium};
        \addlegendentry{$\alpha = 10$}
        \addplot [green, mark=diamond*] table[x=x,y=y0] {\high};
        \addlegendentry{$\alpha = 100$}
    \end{axis}

\end{tikzpicture}
    \vspace{-1em}
    \caption{
      Modularity
    }
    \label{fig:alpha_modularity}
  \end{subfigure}
  \begin{subfigure}[b]{.3\columnwidth}
    \centering
    \begin{tikzpicture}
    \pgfplotstableread{fig/choosing_alpha/lowalpha/modules_per_round.txt}\low
    \pgfplotstableread{fig/choosing_alpha/alpha10/modules_per_round.txt}\medium
    \pgfplotstableread{fig/choosing_alpha/highalpha/modules_per_round.txt}\high
    \pgfplotsset{
        height=3cm+45pt,
        width=.8\textwidth+45pt,
        compat=1.10,
        tick label style={font=\footnotesize},
        xmin=0
    }
    \begin{axis}[
        xlabel=Rounds,
        ymin=0, ymax=10,
        cycle list name=exotic
    ]
        \pgfplotsset{cycle list shift=1}
        \addplot table[x=x,y=y0] {\low};
        \addlegendentry{$\alpha = 1$}
        \addplot [blue, mark=otimes*] table[x=x,y=y0] {\medium};
        \addlegendentry{$\alpha = 10$}
        \addplot [green, mark=diamond*] table[x=x,y=y0] {\high};
        \addlegendentry{$\alpha = 100$}
        \legend{};
    \end{axis}
\end{tikzpicture}
    \vspace{-1em}
    \caption{
      \# Partitions
    }
    \label{fig:alpha_modules}
  \end{subfigure}
  \begin{subfigure}[b]{.3\columnwidth}
    \centering
    \begin{tikzpicture}
    \pgfplotstableread{fig/choosing_alpha/lowalpha/misclassification_per_round.txt}\low
    \pgfplotstableread{fig/choosing_alpha/alpha10/misclassification_per_round.txt}\medium
    \pgfplotstableread{fig/choosing_alpha/highalpha/misclassification_per_round.txt}\high
    \pgfplotsset{
        height=3cm,
        width=.8\textwidth,
        compat=1.10,
        tick label style={font=\footnotesize},
        scale only axis,
        cycle list name=exotic,
        xmin=0,
    }
    \begin{axis}[
        xlabel=Rounds,
        ymin=0, ymax=0.7,
    ]
        \pgfplotsset{cycle list shift=1}
        \addplot table[x=x,y=y0] {\low};
        \addlegendentry{$\alpha = 1$}
        \addplot [blue, mark=otimes*] table[x=x,y=y0] {\medium};
        \addlegendentry{$\alpha = 10$}
        \addplot [green, mark=diamond*] table[x=x,y=y0] {\high};
        \addlegendentry{$\alpha = 100$}
        \legend{};
    \end{axis}
\end{tikzpicture}
    \vspace{-1em}
    \caption{Misclassification}
    \label{fig:alpha_misclassification}
  \end{subfigure}
  \caption{
    Choosing $\alpha$: On the FMNIST-clustered dataset, $\alpha = 10$ strikes the best balance with regards to the three metrics of $G_{clients}$.
  }
  \label{fig:choosingalpha}
\end{figure}

Figure~\ref{fig:choosingalpha} shows the effects of choosing different values of $\alpha$ on the FMNIST-clustered dataset.
Only a subset of 100 clients were included in these experiments, since distinct clusters within $G_{clients}$ can only be observed if the number of nodes in the graph is not continuously increasing.
Nevertheless, the conclusions regarding the choice of $\alpha$ are valid also for experiments including a greater amount of clients.

A low value $\alpha = 1$ leads to decreasing modularity and performs poorly with regards to client similarities within clusters, as can be observed by the high fraction of misclassified clients.
On the other hand, a high value $\alpha = 100$ also shows high and constant modularity, whereas the number of modules can be regarded as too high considering the three clusters in the training data.
The medium value $\alpha = 10$ performs best: modularity is increasing slightly over time, there is a low number of modules and virtually all clients are assigned to a cluster corresponding to their label.

\begin{figure}[h]
  \centering
  \begin{tikzpicture}
  \pgfplotstableread{fig/evaluation/alpha_accuracy/alpha_accuracy-reduced.txt}\data
  \pgfplotsset{
    height=5cm,
    width=\columnwidth,
    compat=1.3,
    legend pos=south east,
  }
  \begin{axis}[
    xlabel=Rounds,
    ylabel=Accuracy,
    ymin= .2, ymax=1,
    cycle list name=exotic,
    legend cell align={left},
    xmin = 0,
    xmax = 100,
  ]
    \addplot+[smooth, thick, mark=triangle] table[x=x,y=y5] {\data};
    \addlegendentry{$\alpha = 0.1$}
    \addplot+[smooth, thick] table[x=x,y=y4] {\data};
    \addlegendentry{$\alpha = 1$}
    \addplot+[smooth, thick, blue, mark=otimes*] table[x=x,y=y3] {\data};
    \addlegendentry{$\alpha = 10$}
    \addplot+[smooth, thick, green, mark=diamond*] table[x=x,y=y2] {\data};
    \addlegendentry{$\alpha = 100$}
  \end{axis}

\end{tikzpicture}
  \vspace{-0.5em}
  \caption{Higher values of $\alpha$ improve the accuracy for the FMNIST-clustered dataset.}
  \vspace{-0.5em}
  \label{fig:alpha_accuracy}
\end{figure}
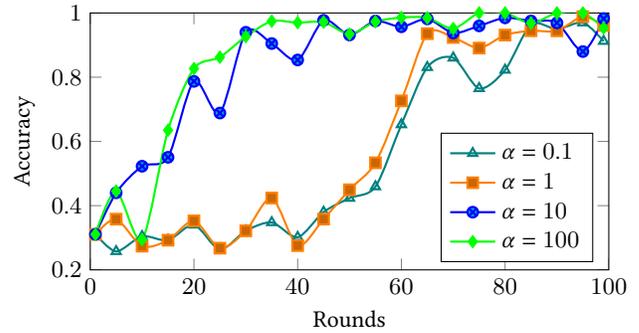

Besides evaluating the impact of of choosing the parameter $\alpha$ on the specialization, we also investigated its impact on the accuracy with the FMNIST-clustered dataset.
Figure~\ref{fig:alpha_accuracy} shows the results for $\alpha$ values between $100$ and $0.1$.

For values of 10 and higher for $\alpha$, the accuracy improves earlier than for values of 1 and lower.
After 100 rounds, the accuracy still comes close to 1 for all values of $\alpha$.
For the lower values of $\alpha$, no specialization emerges in the DAG for this dataset.

The good accuracy of the model after 100 rounds is due to the fact that eventually a generalized model learns to solve the task for all clusters.
In the FMNIST-clustered dataset, the task is simple enough to solve for a generalized model.

While we would expect the accuracy to become worse for very high values of $\alpha$ this doesn't happen for \emph{fully} clustered datasets (i.e. no class overlap between clients from different clusters) as there is no value in generalization.

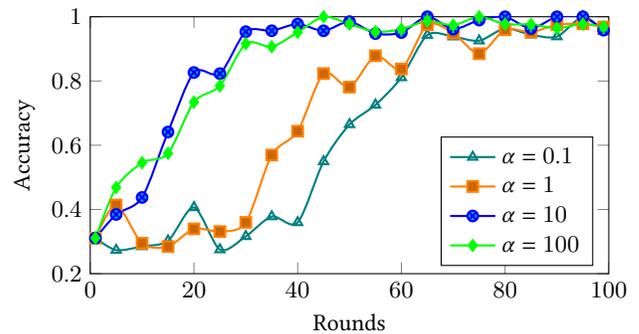
\begin{figure}[h!]
  \centering
  \begin{tikzpicture}
  \pgfplotstableread{fig/evaluation/alpha_accuracy/alpha_accuracy_dynamic-reduced.txt}\data
  \pgfplotsset{
    height=5cm,
    width=\columnwidth,
    compat=1.3,
    legend pos=south east,
  }
  \begin{axis}[
    xlabel=Rounds,
    ylabel=Accuracy,
    ymin= .2, ymax=1,
    cycle list name=exotic,
    legend cell align={left},
    xmin = 0,
    xmax = 100,
  ]
    \addplot+[smooth, thick, mark=triangle] table[x=x,y=y3] {\data};
    \addlegendentry{$\alpha = 0.1$}
    \addplot+[smooth, thick] table[x=x,y=y2] {\data};
    \addlegendentry{$\alpha = 1$}
    \addplot+[smooth, thick, blue, mark=otimes*] table[x=x,y=y1] {\data};
    \addlegendentry{$\alpha = 10$}
    \addplot+[smooth, thick, green, mark=diamond*] table[x=x,y=y0] {\data};
    \addlegendentry{$\alpha = 100$}
  \end{axis}

\end{tikzpicture}
  \vspace{-0.5em}
  \caption{Choosing a dynamic normalization in calculating the weights of models during the tip selection results in better performance for $\alpha = 1$}
  \vspace{-0.5em}
  \label{fig:alpha_accuracy_dynamic}
\end{figure}

The evaluation so far was done using the simple normalization calculation as explained in Section~\ref{sec:approach_specialization}.
Figure~\ref{fig:alpha_accuracy_dynamic} shows the accuracy per round when using the altered normalized accuracy $\mathit{normalized^*}$ to calculate the weights in the random walk.

\begin{figure}[h!]
  \centering
  \begin{tikzpicture}
  \pgfplotstableread{fig/evaluation/alpha_accuracy/alpha_accuracy_relaxed-reduced.txt}\data
  \pgfplotsset{
    height=5cm,
    width=\columnwidth,
    compat=1.3,
    legend pos=south east,
  }
  \begin{axis}[
    xlabel=Rounds,
    ylabel=Accuracy,
    ymin= .2, ymax=1,
    cycle list name=exotic,
    legend cell align={left},
    xmin = 0,
    xmax = 100,
  ]
    \addplot+[smooth, thick, mark=triangle] table[x=x,y=y3] {\data};
    \addlegendentry{$\alpha = 0.1$}
    \addplot+[smooth, thick, orange] table[x=x,y=y2] {\data};
    \addlegendentry{$\alpha = 1$}
    \addplot+[smooth, thick, blue, mark=otimes*] table[x=x,y=y1] {\data};
    \addlegendentry{$\alpha = 10$}
    \addplot+[smooth, thick, green, mark=diamond*] table[x=x,y=y0] {\data};
    \addlegendentry{$\alpha = 100$}
  \end{axis}

\end{tikzpicture}
  \vspace{-0.5em}
  \caption{The effect that better accuracies are achieved faster remains with the relaxed dataset.}
  \vspace{-0.5em}
  \label{fig:alpha_accuracy_relaxed}
\end{figure}
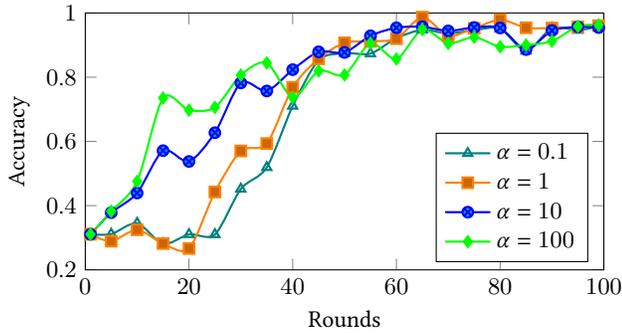

\begin{figure*}[t!]
  \centering
  \newcommand{\alignedincludegraphics}[1]{
    \includegraphics
    [trim={1.32cm 0 6cm 1.5cm},
    clip,
    align=c,
    width=0.295\linewidth, page=1]
    {fig/evaluation/accuracy_table/#1}}
  \begin{tabular}{cccc}
    & FMNIST-clustered & Poets & CIFAR-100 \\
    \rotatebox[origin=c]{90}{FedAvg} &
    \alignedincludegraphics{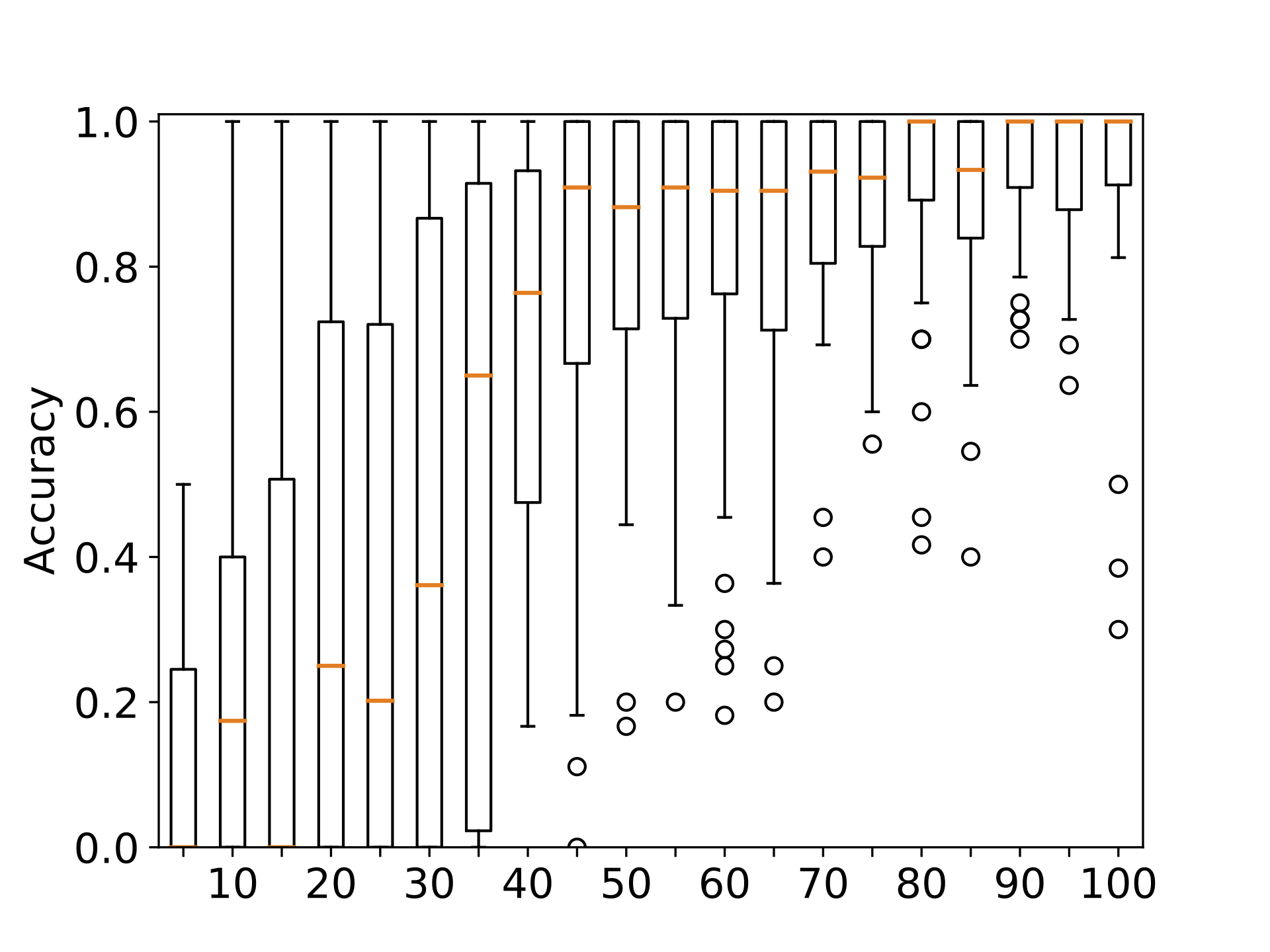} &
    \alignedincludegraphics{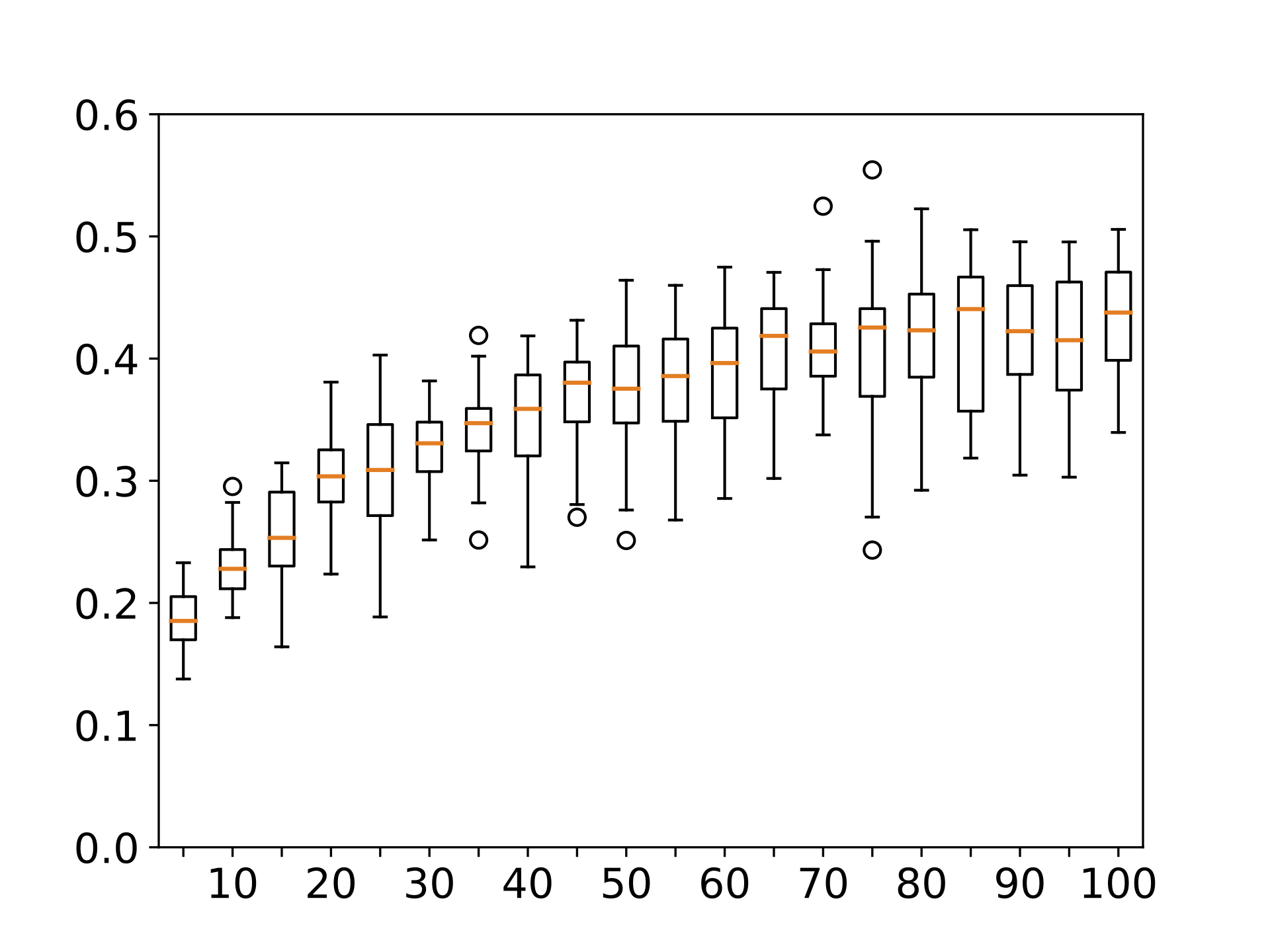} &
    \alignedincludegraphics{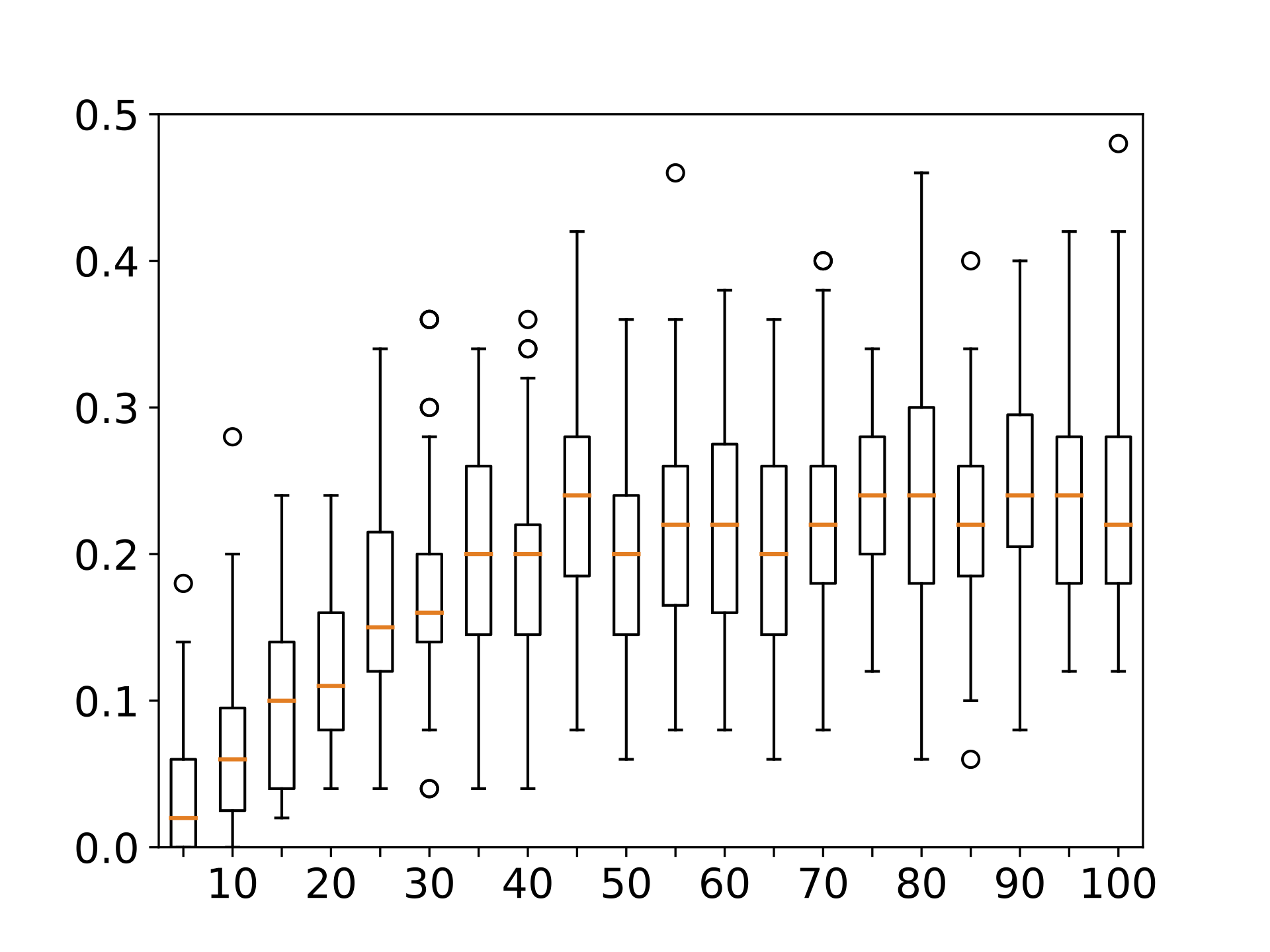} \\ 
    \rotatebox[origin=c]{90}{Specializing DAG}  &
    \alignedincludegraphics{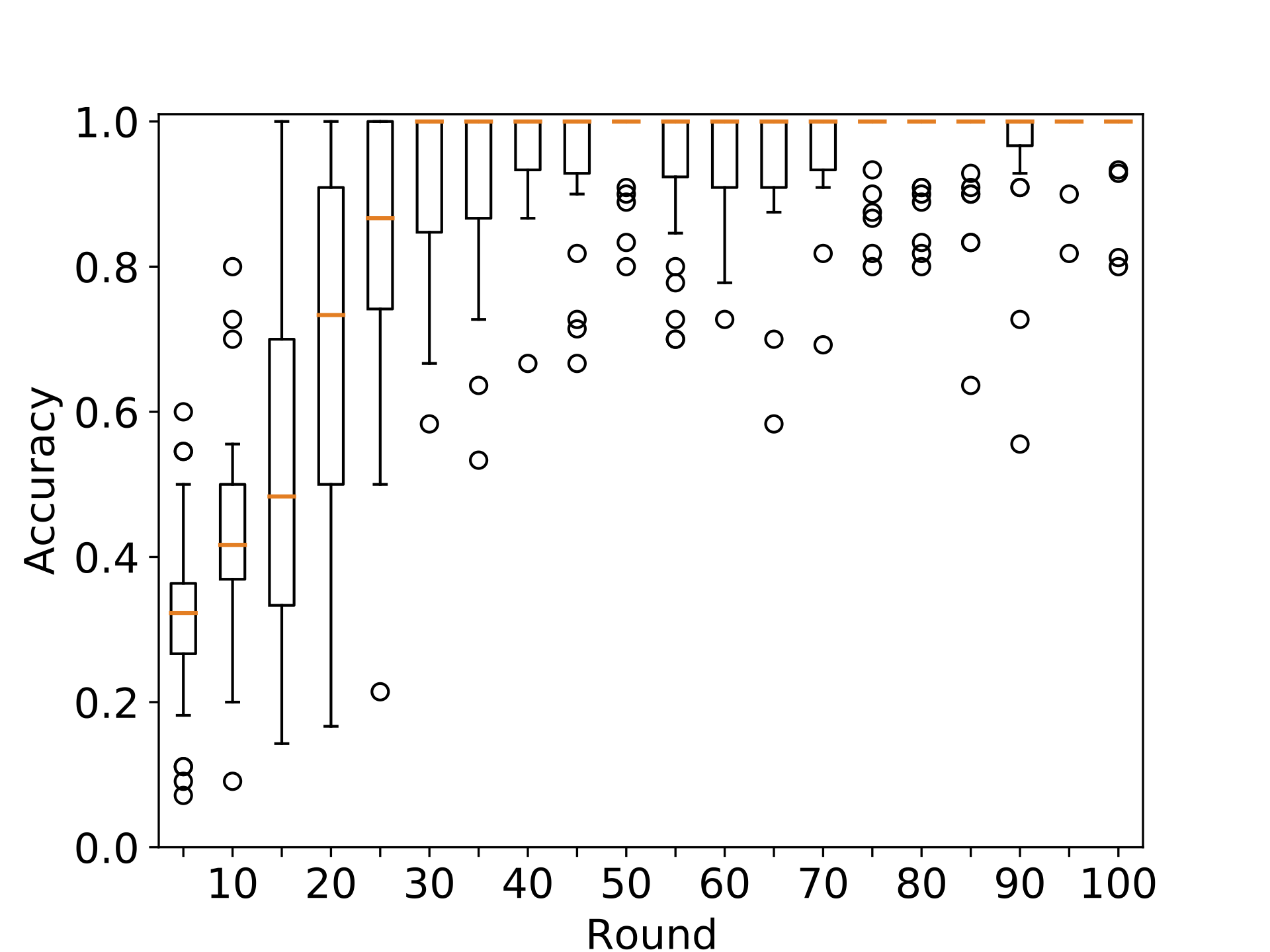} & %
    \alignedincludegraphics{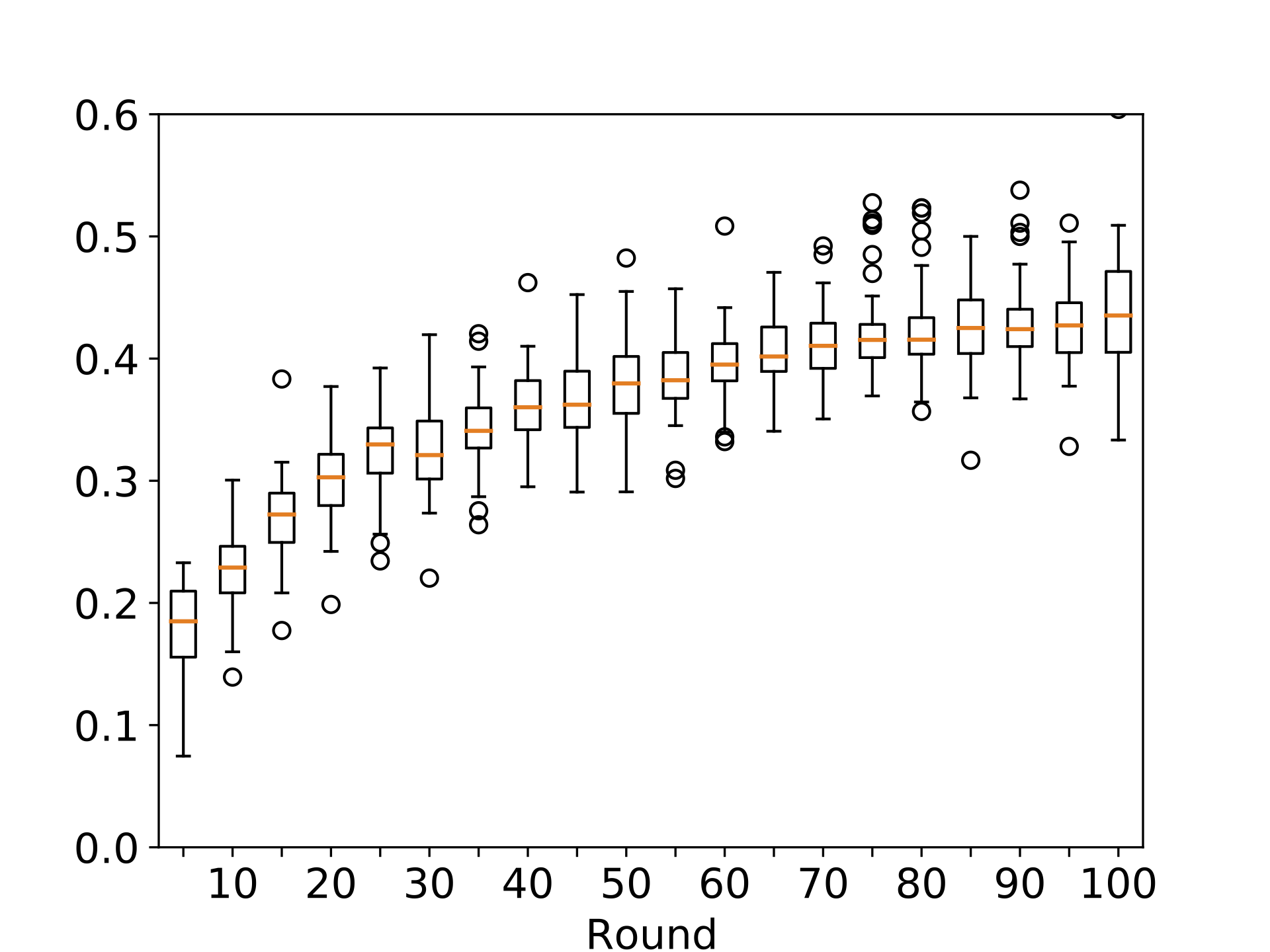} & %
    \alignedincludegraphics{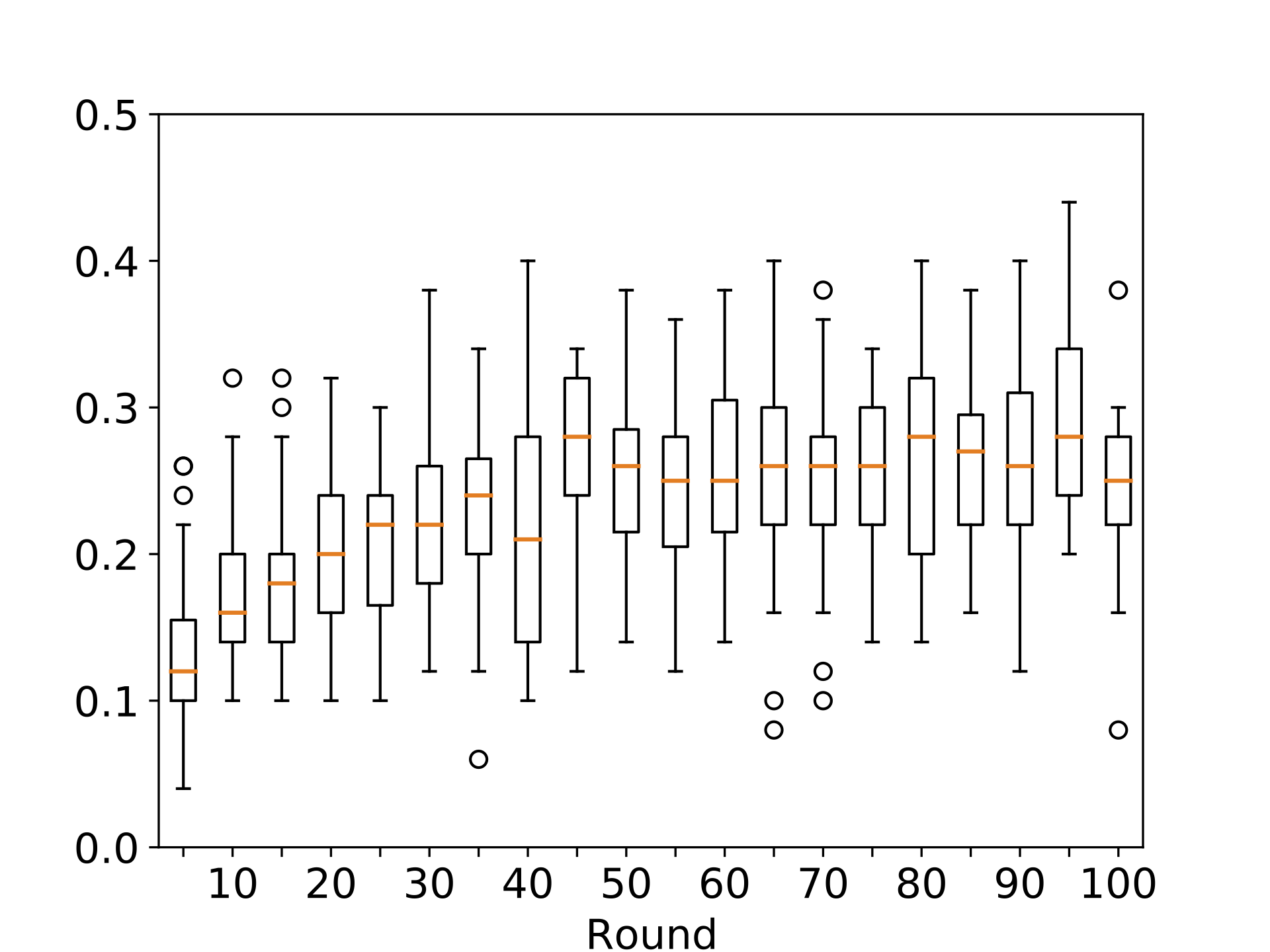} \\ %
  \end{tabular}
  \caption{The accuracy on client-local data compared between federated averaging (FedAvg) and our approach (Specializing DAG), grouped over 5 rounds. FedAvg uses a central averaged model, while the DAG uses the specialized local models.}
  \label{fig:accuracy_table}
\end{figure*}

The usage of this altered normalization improves the accuracy slightly for $\alpha = 1$.
This corresponds to a higher approval pureness of 0.51 for $\alpha = 1$ when using the dynamic normalization, compared to $0.40$ with the standard normalization.

To evaluate the performance of our approach on not fully clustered data we created a relaxed FMNIST-clustered dataset, where each cluster contains between 15 and 20 percent of data from other clusters.
Figure~\ref{fig:alpha_accuracy_relaxed} shows the accuracy for different values of $\alpha$ for this relaxed dataset.

In general the relaxation of the clusters helps the model to generalize faster, resulting in better performance even for low values of $\alpha$.
At the same time, the performance of the well specialized cases with high values of $\alpha$ improve slightly slower because of the relaxation.
Thus, while the same effect remains in this dataset, it is weaker than in the fully clustered dataset.

\subsubsection{Comparison with Federated Averaging}
We compare our approach with federated averaging as a baseline using the three different datasets.
We first present the approval pureness of our approach for each dataset and then discuss the accuracy results.

\begin{table}[h!]
  \centering
  \begin{tabular}{cccc}
    Dataset & \# clusters & base pureness & pureness \\ \hline
    FMNIST-clustered & 3 & 0.33 & 1.0\\
    Poets & 2 & 0.5 & 0.95\\
    CIFAR-100 & 20 & 0.05 & 0.51
  \end{tabular}
  \caption{The approval pureness in the DAG after 100 rounds of training with our approach.}
  \label{table:pureness}
\end{table}

To quantify how strong the DAG specialized in these experiments, we show the approval pureness in Table~\ref{table:pureness}.
In the FMNIST-clustered dataset there are three clusters for the groups of classes, Poets has the two clusters for texts by Goethe and Shakespeare and CIFAR-100 is clustered into the 20 superclasses.
The base pureness is the approval pureness expected if the approvals would be randomly spread over all clusters.

The approvals in the DAG show perfect pureness for the FMNIST-clustered dataset.
That is, all approvals of models are from within the same cluster, which is sensible for a completely clustered dataset where the integration of models from other clusters will not lead to better performance.
For Poets and CIFAR-100, the approval pureness is lower and shows the balance between generalization and specialization into the clusters.

Figure~\ref{fig:accuracy_table} shows an overview of the accuracy of our approach compared to federated averaging.
The values show the accuracy distribution on the local data of all clients selected in five consecutive rounds using the aggregated model in FedAvg and the locally optimized and published model for the Specializing DAG.

The accuracy evaluation shows that our approach performs better for the FMNIST-clustered dataset, where the accuracy improves faster.
The larger deviation in federated averaging shows the missing ability to specialize.
Consequently, the DAG is the first mechanism that enables training of a machine learning model on heterogeneous client datasets in a decentralized and asynchronous way.

For the Poets and CIFAR-100 datasets, the DAG achieves similar accuracy results compared to federated averaging.
This also shows the feasibility of the decentralized approach: the central server can be removed without an accuracy penalty for the evaluated datasets.
Further improving the training accuracy on these datasets is an area of future work.

\subsubsection{Comparison with FedProx}

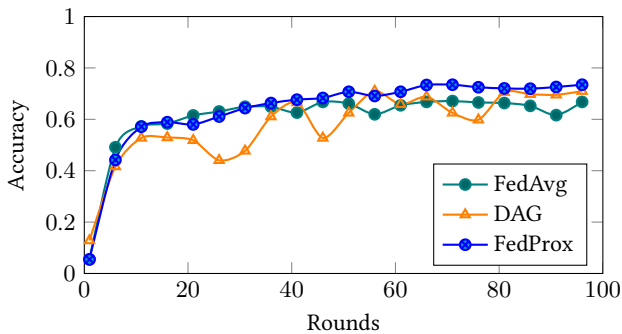
\begin{figure}[b]
  \centering
  \begin{tikzpicture}
  \pgfplotstableread{fig/evaluation/fedprox/accuracy.txt}\data
  \pgfplotsset{
    height=5cm,
    width=\columnwidth,
    compat=1.3,
    legend pos=south east,
  }
  \begin{axis}[
    xlabel=Rounds,
    ylabel=Accuracy,
    ymin= 0, ymax=1,
    cycle list name=exotic,
    legend cell align={left},
    xmin = 0,
    xmax = 100,
  ]
    \addplot+[smooth, thick] table[x=x,y=fedavg] {\data};
    \addlegendentry{FedAvg}
    \addplot+[smooth, thick, mark=triangle] table[x=x,y=tangle] {\data};
    \addlegendentry{DAG}
    \addplot+[smooth, thick, blue, mark=otimes*] table[x=x,y=fedprox] {\data};
    \addlegendentry{FedProx}
  \end{axis}

\end{tikzpicture}
  \caption{
    Initially, the average accuracy is more consistent using the centralized approaches; later, the DAG performance stabilizes and outperforms FedAvg.
  }
  \label{fig:fedprox_accuracy}
\end{figure}

\begin{figure}[t]
  \centering
  \begin{tikzpicture}
  \pgfplotstableread{fig/evaluation/fedprox/loss.txt}\data
  \pgfplotsset{
    height=5cm,
    width=\columnwidth,
    compat=1.3,
    legend pos=north east,
  }
  \begin{axis}[
    xlabel=Rounds,
    ylabel=Loss,
    ymin= 0, ymax=4,
    cycle list name=exotic,
    legend cell align={left},
    xmin = 0,
    xmax = 100,
  ]
    \addplot+[smooth, thick] table[x=x,y=fedavg] {\data};
    \addlegendentry{FedAvg}
    \addplot+[smooth, thick, mark=triangle] table[x=x,y=tangle] {\data};
    \addlegendentry{DAG}
    \addplot+[smooth, thick, blue, mark=otimes*] table[x=x,y=fedprox] {\data};
    \addlegendentry{FedProx}
  \end{axis}

\end{tikzpicture}
  \caption{
    The DAG consistently performs better with regards to the average loss compared to FedAvg.
    A small margin remains compared to the centralized FedProx approach.
  }
  \label{fig:fedprox_loss}
\end{figure}
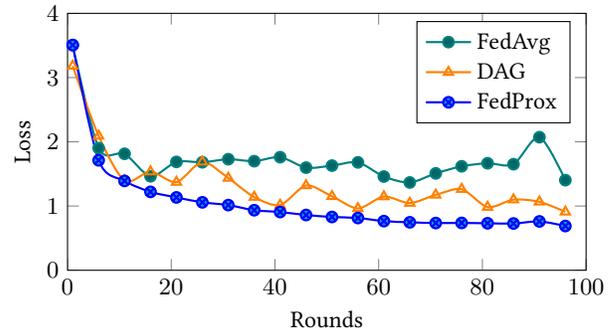

FedProx \cite{liFederatedOptimizationHeterogeneous2020} is a state-of-the-art extension of FedAvg that guarantees model convergence even with non-IID client data distributions.
The authors claim that in realistic scenarios, FedAvg only receives partial information from the clients, which it does not properly account for.
These partial information can stem from statistical heterogeneity (non-IID data distributions) as well as stragglers, i.e. clients that were only able to submit partially trained models in time.
Thus, the authors propose to add a proximal term to FedAvg that improves the convergence behavior in such heterogeneous networks theoretically and empirically.

In the case of the specializing DAG, there are no stragglers due to its asynchronous nature:
In a distributed implementation, each client continuously runs the training process as often as its resources permit, independent from all other clients.
We only introduce the concept of rounds to be able to compare the performance of the DAG with centralized approaches.

For comparing ourselves with FedProx and unmodified FedAvg, we used the synthetic dataset proposed by FedProx:
It is parameterized with $\alpha = 0.5$, $\beta = 0.5$, $\alpha, \beta \in [0;1]$, where $\alpha$ and $\beta$ control the dissimilarity of the local training samples for each client and between clients, respectively.

Figures \ref{fig:fedprox_accuracy} and \ref{fig:fedprox_loss} show the average accuracy and loss in a scenario with 30 clients in total and 10 active clients per round.
The variance in accuracy and loss of the DAG is generally higher compared to the centralized approaches which can be explained by the statistical tip selection process as part of training and inference.
However, the Specializing DAG eventually outperforms FedAvg in both accuracy and loss without the need for a central parameter server.
Regarding the loss, the DAG results come close to the FedProx baseline, which shows how implicit specialization in the DAG effectively helps to accommodate differences in the data distribution among clients.

\subsubsection{Poisoning}

In order to investigate the robustness of the approach, we conducted experiments with flipped-label poisoning attacks using the original FMNIST dataset that is split by the authors of the handwritten digits.
Specifically, we exchanged the labels 3 and 8 for a subset of clients after 100 training rounds without any data poisoning.

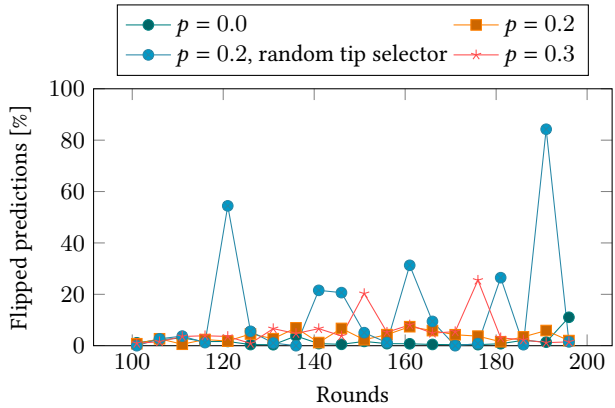
\begin{figure}[t]
  \centering
  \begin{tikzpicture}
    \pgfplotstableread{fig/poisoning/accuracyts-20/average_label_flip_misclassification_per_round.txt}\accuracytslow
    \pgfplotstableread{fig/poisoning/defaultts-20/average_label_flip_misclassification_per_round.txt}\defaulttslow
    \pgfplotstableread{fig/poisoning/defaultts-all/average_label_flip_misclassification_per_round.txt}\defaulttsall
    \pgfplotstableread{fig/poisoning/accuracyts-30/average_label_flip_misclassification_per_round.txt}\accuracytshigh
    \pgfplotstableread{fig/poisoning/nopoisoning/average_label_flip_misclassification_per_round.txt}\nopoisoning
    \pgfplotsset{
        height=5cm,
        width=\columnwidth,
        compat=1.3,
        legend pos=north west,
    }
    \begin{axis}[
        xlabel=Rounds,
        ylabel={Flipped predictions [\%]},
        ymin=0, ymax=100,
        cycle list name=exotic,
        legend cell align={left},
        legend columns=2,
        legend style={at={(0.5,1.05)}, anchor=south},
    ]
        \addplot table[x=x,y=y0] {\nopoisoning};
        \addlegendentry{$p=0.0$}
        \addplot table[x=x,y=y0] {\accuracytslow};
        \addlegendentry{$p=0.2$}
        \addplot table[x=x,y=y0] {\defaulttsall};
        \addlegendentry{$p=0.2$, random tip selector}
        \addplot table[x=x,y=y0] {\accuracytshigh};
        \addlegendentry{$p=0.3$}
    \end{axis}
\end{tikzpicture}
  \caption{Flipped predictions of samples in the classes 3 and 8.}
  \label{fig:labelflip_mispredictions}
\end{figure}

Figure~\ref{fig:labelflip_mispredictions} illustrates the success of the poisoning attack in different scenarios:
It shows how many samples of the classes 3 and 8 in the clients' test datasets were mispredicted as belonging to the other class using the reference model that the clients selected from the DAG.
The parameter $p$ defines the fraction of clients that used poisoned training and test data.
As a baseline, we also measured the behavior of the original, purely random tip selector.

Compared to the baseline results, the effects of the attack with $p = 0.2$ on the overall network are very limited and almost within the variance that is also present with $p = 0.0$.
When increasing the number of poisoned clients to $p = 0.3$, the effects are noticeable, but still below 30\% mispredictions overall.

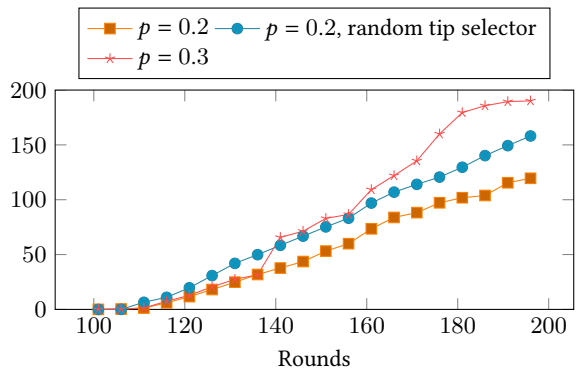
\begin{figure}[t]
  \centering
  \begin{tikzpicture}
        \pgfplotstableread{fig/poisoning/accuracyts-20/average_num_approved_poisoned_transactions_in_consensus_per_round.txt}\accuracytslow
        \pgfplotstableread{fig/poisoning/defaultts-20/average_num_approved_poisoned_transactions_in_consensus_per_round.txt}\defaulttslow
        \pgfplotstableread{fig/poisoning/defaultts-all/average_num_approved_poisoned_transactions_in_consensus_per_round.txt}\defaulttsall
        \pgfplotstableread{fig/poisoning/accuracyts-30/average_num_approved_poisoned_transactions_in_consensus_per_round.txt}\accuracytshigh
        \pgfplotstableread{fig/poisoning/nopoisoning/average_num_approved_poisoned_transactions_in_consensus_per_round.txt}\nopoisoning
    \pgfplotsset{
        height=4.5cm,
        width=\columnwidth,
        compat=1.3,
        legend pos=north west,
    }
    \begin{axis}[
        xlabel=Rounds,
        ymin=0, ymax=200,
        cycle list name=exotic,
        legend cell align={left},
        legend columns=2,
        legend style={at={(0.5,1.05)}, anchor=south},
    ]
        \pgfplotsset{cycle list shift=1}
        \addplot table[x=x,y=y0] {\accuracytslow};
        \addlegendentry{$p=0.2$}
        \addplot table[x=x,y=y0] {\defaulttsall};
        \addlegendentry{$p=0.2$, random tip selector}
        \addplot table[x=x,y=y0] {\accuracytshigh};
        \addlegendentry{$p=0.3$}
    \end{axis}
\end{tikzpicture}
  \caption{
    Average number of approved poisonous transactions in the consensus.
  }
  \label{fig:labelflip_approved_transactions}
\end{figure}

Looking at the selected reference transactions in more detail, we can observe that, although the poisoning did not have severe effects, a high number of poisoned updates are included in the reference transactions by direct or indirect approvals.
Especially noteworthy is that the poisoning impact on the mispredictions is higher for the random tip selector with $p = 0.2$ than for the accuracy tip selector with $p = 0.3$, even though the number of approved poisoned transactions is lower for the former.

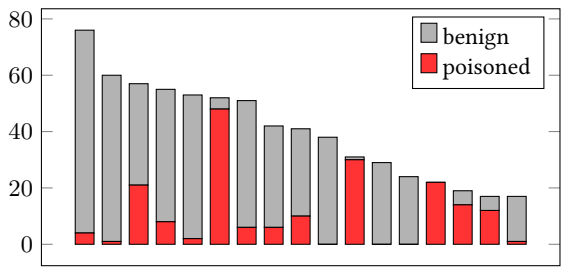
\begin{figure}[b]
  \centering
  \begin{tikzpicture}
    \pgfplotstableread{fig/poisoning/clusters.txt}\clusters
    \pgfplotsset{
        height=5cm,
        width=\columnwidth,
        legend pos=north west,
    }
    \begin{axis}[
        ybar stacked,
        xtick=data,
        legend style={cells={anchor=west}, legend pos=north east},
        reverse legend=true, %
        xticklabels from table={\clusters}{i},
        xticklabel style={text width=1cm,align=center},
        xmajorticks=false,
        bar width=0.25cm,
    ]
        \addplot [fill=red!80] table [y=x, meta=i, x expr=\coordindex] {\clusters};
        \addlegendentry{poisoned}
        \addplot [fill=gray!60] table [y=y, meta=i, x expr=\coordindex] {\clusters};
        \addlegendentry{benign}
    \end{axis}
\end{tikzpicture}
  \caption{Distribution of poisoned clients over 15 inferred clusters for $p = 0.3$.}
  \label{fig:poisoned_clusters}
\end{figure}

This can be explained by the containment of poisoned transactions within a subset of clients.
While a poisoned transaction may be incorporated into another clusters' consensus from time to time, leading to a high number of indirectly approved poisoned transactions, in most cases it is other malicious clients that approve a poisoned transaction:

Figure~\ref{fig:poisoned_clusters} depicts the distribution of poisoned clients over the clusters reconstructed by the Louvain algorithm.
Most of them end up in clusters where a majority of other clients are also affected by the attack.

While this protects other network participants, it also means that the attack is difficult to detect for the affected clients.
If the goal of the attack is known in advance, clients could use the random tip selector to obtain a reference transaction that is most likely not affected by the attack in order to cross-check their locally trained model.

\subsubsection{Scalability}

Compared to gossiping approaches for federated learning, the main overhead of the proposed DAG consists of the time that each client needs to perform the random walk and to evaluate models on the local data as part of it.
If the time needed for the random walk increases with the number of updates in the DAG or the number of participating clients, this would limit the scalability of the approach.
This section evaluates the time needed for the random walk when training the original FMNIST dataset with increasing numbers of clients that are concurrently performing model training.

\begin{figure}
  \centering
  \begin{tikzpicture}
  \pgfplotstableread{fig/evaluation/scalability/duration.txt}\data
  \pgfplotsset{
    height=5cm,
    width=\columnwidth,
    compat=1.3,
    legend pos=north east,
  }
  \begin{axis}[
    xlabel=Rounds,
    ylabel={Random Walk Duration [s]},
    cycle list name=exotic,
    legend cell align={left},
    xmin = 0,
    xmax = 100,
    legend cell align={left},
    legend columns=2,
    legend style={at={(0.5,1.05)}, anchor=south},
  ]
    \addplot+[smooth, thick] table[x=x,y=clients5] {\data};
    \addlegendentry{5 active clients}
    \addplot+[smooth, thick, mark=triangle] table[x=x,y=clients10] {\data};
    \addlegendentry{10 active clients}
    \addplot+[smooth, thick, blue, mark=otimes*] table[x=x,y=clients20] {\data};
    \addlegendentry{20 active clients}
    \addplot+[smooth, thick, green, mark=diamond*] table[x=x,y=clients40] {\data};
    \addlegendentry{40 active clients}
  \end{axis}

\end{tikzpicture}
  \caption{
    Development of time required for the random walk for different numbers of concurrently active clients over the course of 100 training rounds.
    The differences for increasing numbers of active clients are marginal which indicates a good scalability of the approach.
  }
  \label{fig:scalability}
\end{figure}
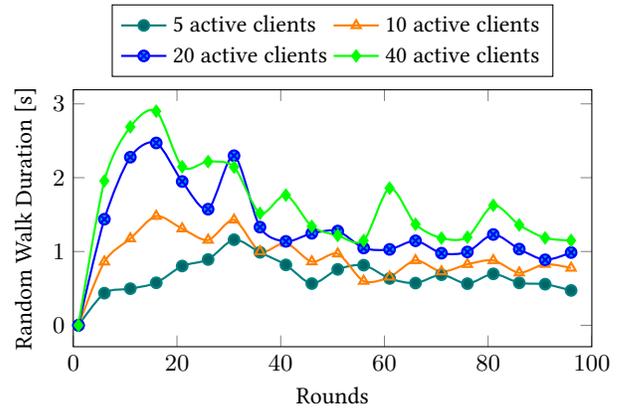

Generally, the random walk makes up a significant amount of the required compute resources compared to model training:
In our example, training the FMNIST model takes about 300ms whereas the time required for the random walk ranges from 600-1200ms (cf. Figure \ref{fig:scalability}).
However, in a real-world implementation, the time required for the random walk can be hidden between training runs, since it would be sensible for a client to only perform training in set intervals or when new training data arrives.

In our scalability experiments, we started the random walk at a transaction sampled at a depth of 15-25 transactions from the tips, as proposed by Popov \cite{sergueipopov_tangle_2017}.

Figure \ref{fig:scalability} shows the average time that a single client spends on the random walk, for increasing numbers of clients that are training concurrently.
The concurrency in the network has an impact on the random walk because well- performing models will generally have a greater number of direct child transactions that were created simultaneously and all of which have to be evaluated on local data during the random walk.

Especially in the early phases of the collaborative training process, this number is not well balanced among the transactions since there are still large differences in accuracy between them.
However, as the trained models improve, the variance in the number of child transaction levels out.
In practice, this would also require ideal network conditions, i.e. all new transactions are broadcasted equally well among network participants.

In conclusion, the concurrency in the network has little impact on the costs incurred by the DAG algorithm, and hence it can be expected to scale well also for larger numbers of clients.

\section{Conclusion and Future Work} %
\label{sec:conclusion_and_future_work}

We presented a novel approach to achieve specialized models in federated learning without a central server: by using a DAG for the communication of models and an accuracy-biased random walk, we show the manifestation of clusters of clients with similar local data.
This specialization emerges directly from the way the DAG is used to communicate model updates, thus creating a unified solution for decentralized and personalized federated learning.
We enable a tradeoff between reaching a consensus on a generalized model and specializing (personalizing) the models to local data.
We evaluated this approach with our prototypical implementation on three datasets and showed equal or better learning performance in a simulation.
Finally, we showed the poisoning resistance and scalability of our approach.

In the future we would like to integrate ideas from multi-task and personalized federated learning such as training only some layers of the machine learning model.

\section*{Acknowledgments}
We would like to thank the anonymous reviewers for their helpful comments on earlier versions of this paper. 
We are also thankful for the comments by Felix Eberhardt and Daniel Richter from our research group.

This research was partly funded by the Federal Ministry for Economic Affairs and Energy of Germany as part of the program ”Smart Data” (project number 01MD19014C),  by the German Federal Ministry of Transport and Digital Infrastructure through the mFUND (project number 19F2093C) and by the Federal Ministry of Education and Research of Germany in the framework of KI-LAB-ITSE (project number 01IS19066).

\bibliographystyle{ACM-Reference-Format}
\bibliography{paper.bib, additional.bib}
\end{document}